\documentclass[hidelinks,onefignum,onetabnum]{siamart251104}
\ifpdf
\hypersetup{
  pdftitle={Bessel Functions and Analysis of Circular Waveguides},
  pdfauthor={J. Mora-Paz, L. Demkowicz, C. G. Taylor, J. Grosek, S. Henneking}
}
\fi
\usepackage{amsmath}
\usepackage{amsbsy}
\usepackage{amsfonts}
\usepackage{amssymb}
\usepackage{array}
\usepackage{bm}
\usepackage{booktabs}
\usepackage{comment}
\usepackage{csquotes}
\usepackage{enumerate}
\usepackage{float}
\usepackage{mathrsfs}
\usepackage{pgf}
\usepackage{mathtools}
\usepackage{derivative}
\usepackage[font={footnotesize}]{caption}
\usepackage{subcaption}
\usepackage{soul}
\usepackage{multirow}
\usepackage{algorithm}
\usepackage[noend]{algpseudocode}
\usepackage{placeins} 

\renewcommand{\theequation}{\thesection.\arabic{equation}}
\def\be{\begin{equation}}
\def\ee{\end{equation}}
\def\ba{\begin{array}}
\def\ea{\end{array}}
\def\bea{\begin{eqnarray}}
\def\eea{\end{eqnarray}}
\def\beas{\begin{eqnarray*}}
\def\eeas{\end{eqnarray*}}

\newcommand{\bb}[1]{\mathbb{#1}}

\newcommand{\bfJ}{\boldsymbol{J}}

\newcommand{\bfR}{\boldsymbol{R}}
\newcommand{\bfS}{\boldsymbol{S}}

\newcommand{\ptl}{{\partial}}

\newcommand{\doubleIC}{\mathbb{C}}

\newcommand{\ds}{\displaystyle}

\newcommand{\defeq}{\coloneqq}

\newcommand{\mapsfrom}{\mathrel{\reflectbox{\ensuremath{\mapsto}}}}

\newcommand{\p}{\partial}
\newcommand{\pp}[2]{\frac{\partial #1}{\partial #2}}

\newcommand{\dd}[2]{\frac{d #1}{d #2}}
\newcommand{\ddn}[3]{\frac{d^{#1} #2}{d#3^{#1}}}

\newcommand{\besseltol}{\varepsilon_\text{trunc}}
\newcommand{\eigentol}{\varepsilon_\text{eig}}

\definecolor{deepskyblue}{RGB}{0, 191, 255}
\definecolor{blueviolet}{RGB}{138, 43, 226}

\newcommand{\hlgray}[1]{\colorbox{gray!50}{#1}}

\definecolor{dgreen}{rgb}{0.2, 0.5, 0.2}

\catcode `@=11

\newbox \itemlist@label
\newdimen \itemlist@labelpad

\def\itemlist@makelabel#1{%
\setbox\itemlist@label =\hbox{#1}%
\ifdim \wd\itemlist@label >\labelwidth
\itemlist@labelpad=\textwidth
\advance\itemlist@labelpad by -\rightmargin
\advance\itemlist@labelpad by -\@totalleftmargin
\advance\itemlist@labelpad by \labelwidth
\hbox to \itemlist@labelpad {#1\hfill}%
\else #1\hfill
\fi
}

\renewcommand{\Re}{\operatorname*{Re}}
\renewcommand{\Im}{\operatorname*{Im}}

\let\tilde\widetilde

\newcommand{\nml}{\mathbf{n}}
\newcommand{\wavenum}{\kappa}
\newcommand{\mustraight}{\mu_{\text{straight}}}
\newcommand{\core}{\text{core}}
\newcommand{\clad}{\text{clad}}
\newcommand{\lclad}{\text{left}}
\newcommand{\rclad}{\text{right}}
\newcommand{\Bfirst}{V}
\newcommand{\Bsecond}{W}
\newcommand{\zpml}{z_{\text{PML}}}
\newcommand{\pmlparameter}{C_{\text{PML}}}

\newsiamremark{remark}{Remark}
\newsiamremark{hypothesis}{Hypothesis}
\crefname{hypothesis}{Hypothesis}{Hypotheses}
\newsiamthm{claim}{Claim}
\newsiamremark{fact}{Fact}
\crefname{fact}{Fact}{Facts}

\newcommand{\revA}[1]{#1}
\newcommand{\revB}[1]{#1}
\newcommand{\revC}[1]{#1}

\graphicspath{{figures/}}
\headers{Bessel Functions and Analysis of Circular Waveguides}{J. Mora-Paz, L. Demkowicz, C. G. Taylor, J. Grosek, S. Henneking}
\title{Bessel Functions and Analysis of\\ Circular Waveguides
\thanks{
\textbf{Funding:} Authors 1, 2 and 5 were supported by AFOSR grant FA9550-23-1-0103.
}
}
\author{Jaime Mora-Paz\thanks{Corresponding author. Oden Institute for Computational Engineering and Sciences, The University of Texas at Austin, Austin, TX 78712, USA
  (\email{jdmorap@utexas.edu}).} 
\and 
Leszek Demkowicz\thanks{Oden Institute for Computational Engineering and Sciences, The University of Texas at Austin, Austin, TX 78712, USA
  (\email{leszek@oden.utexas.edu}, \email{christina.taylor@austin.utexas.edu}, \email{\mbox{stefan@oden.utexas.edu}}).}
\and 
Christina G. Taylor\footnotemark[3]
\and\newline
Jacob Grosek\thanks{Directed Energy Directorate, Air Force Research Laboratory, Albuquerque, NM 87117, USA
  (\email{jacob.grosek.1@us.af.mil}).}
\and 
Stefan Henneking\footnotemark[3]
}
\begin{document}
\maketitle
\begin{abstract}
This paper is devoted to the study of circularly coiled optical slab waveguides, which is also applicable to acoustical waveguides.
We use a change of variables and the classical Frobenius method to compute Bessel functions of complex order and complex argument, and combine it with a perfectly matched layer technique to solve the relevant Bessel eigenvalue problem and deliver accurate loss factors for eigensolutions to the three-layer optical slab waveguide problem. 
The solutions provide a benchmark for verifying model implementations of this problem and allow for a numerical verification of the Glazman criterion that provides a foundation for the well-posedness and stability analysis of homogeneous circular waveguides with impedance boundary conditions.
\end{abstract}
%
%
%
\section{Introduction}
\label{sec:introduction}

Fiber optic waveguides play important roles in many photonic applications, including sensing, scientific experimentation, telecommunications, healthcare, defense technologies, and others.
In order to satisfy system spatial restrictions (compact packaging), to control the degree of guidance loss (i.e., differential mode bend loss), and/or to ensure efficient heat dissipation onto a thermally conductive spool/mandrel, these waveguides are typically coiled~\cite{taylor1984bending, richardson2010high, power2009mccomb}. 
Locally, in the coiled regions of the waveguide, the bending can be well approximated \revC{as circular}, with a known bend radius\revA{, denoted throughout this manuscript by $r_0$}. 
This circular bending affects the propagation of the light through the optical waveguide, especially via mode bend losses. 
\revC{
However, accurately predicting these bend losses in computational models remains challenging, in part because establishing a mathematically well-posed formulation of the three-dimensional electromagnetic waveguide problem with appropriate boundary conditions (BCs) is itself nontrivial.
}

The present work started as part of an initiative to perform a computational verification of our three-dimensional circularly bent fiber finite element model's methodology \cite{mora2026bent}, implemented in $hp$3D~\cite{henneking2024hp3d}, based on a vectorial envelope formulation~\cite{henneking2025envelope, henneking2026comparison} of a time-harmonic Maxwell fiber model~\cite{henneking2021fiber, henneking2022parallel}. 
To do so, we consider a simplified, circularly bent, two-dimensional (slab) waveguide that has \revC{smooth, separable} solutions (derived and computed below) \revC{that depend on $r_0$}.
\revA{As shown below, the resulting mathematical model is an eigenvalue problem for Bessel's differential equation subject to appropriate boundary, interface, and scaling conditions. The eigenvalue determines the order of the corresponding Bessel solutions, while the eigenfunction is, up to a scaling constant, a linear combination of two fundamental solutions of the differential equation (e.g., two Bessel functions or two Hankel functions of that order). We solve this eigenproblem through a nonlinear system whose complex-valued unknowns are the eigenvalue and the coefficients of this linear combination, with the governing equations arising from the prescribed boundary, interface, and scaling conditions.}

The aim of this paper is to develop a trustworthy procedure to solve the Bessel eigenproblem \revC{(described in detail below) for two} different purposes:
\begin{itemize}
    \item Provide a high-accuracy verification for our 3D finite element implementation for modeling light propagation in bent optical waveguides.
    \item \revC{Determine the loss factors of the propagating modes of a three-layer circular waveguide using realistic data corresponding to step-index optical fibers, characterize their dependence on the bend radius $r_0$, and compare them with results available in the literature \cite{marcatili1969bends, takuma1981bent, marcuse1982light}.}
\end{itemize}

\revA{The main computational challenges arise from the ranges of the physically relevant parameters in our problem. In particular, both the order and the argument of the relevant Bessel functions are complex-valued and have magnitudes in the $10^5$--$10^7$ range. Moreover, the computations require very high sensitivity in order to accurately resolve the small perturbations in the eigenvalues as $r_0$ varies. The corresponding perturbations in the imaginary parts of the eigenvalues may be tens of orders of magnitude smaller than those in the real parts. Consequently, solving the eigenvalue problem requires the ability to evaluate Bessel functions of large complex order at complex arguments both rapidly and reliably using extended-precision arithmetic.}

\revA{The central premise of the present work is that, for the parameter regime of interest, developing custom procedures for evaluating solutions to Bessel's equation enables the desired functions to be evaluated efficiently and to arbitrarily high precision, whereas such evaluations can be challenging with standard software packages for Bessel and Hankel functions. To illustrate the difficulty of evaluating Bessel functions of large complex order and large argument, we invite the reader to plot $J_\beta(\wavenum r)$ (i.e., the Bessel function of the first kind of order $\beta$ evaluated at $\wavenum r$) for $r\in[r_0-b,r_0+b]$, with $r_0=10^4$, $b=1$, $\wavenum=10^2$, and $\beta=\sqrt{10^{12}-100i}$, using the mathematical software of their choice, and to assess both the computational cost and the sensitivity of the result to small perturbations in $\beta$.\footnote{The values suggested here are representative of the physically relevant parameter ranges considered in the numerical experiments below.}}

\revA{Concerning the second purpose of our Bessel eigensolver, several authors in optics have avoided the computational difficulties described above because their objective was not to resolve the eigenvalues and eigenfunctions in detail, but rather to obtain closed-form approximations of bend losses in waveguides and optical fibers. To this end, they employ combinations of analytical approximations and asymptotic techniques that lead to explicit loss formulas.}

\revA{One of the earliest such results is due to Marcatili~\cite{marcatili1969bends}. His analysis applies to circular waveguides with rectangular cross-sections and allows for four different refractive indices across the subdomains; our slab-waveguide problem corresponds to a special case requiring only two indices. The derivation of Marcatili's loss formula involves solving an approximate complex-valued eigenvalue equation, making extensive use of properties of Hankel and Bessel functions, and introducing approximations based on the large bending ratio and the near-unity ratio of the refractive indices. As a second reference, we consider the classic book by Marcuse~\cite{marcuse1982light}. The corresponding loss estimate avoids solving directly for a complex-valued eigenvalue and instead makes use of the ``approximation by tangents'' for Hankel functions given in~\cite[8.452]{gradshteyn2014table}, together with additional algebraic manipulations and asymptotic approximations. A third relevant reference is the work of Takuma et al.~\cite{takuma1981bent}, whose approach is based on a change of variables and a truncated expansion in terms of the bend radius, combined with solutions of the straight slab-waveguide eigenvalue equation and additional approximations and corrections.}

\revC{Because of the physics of the problem under consideration, comparing the loss estimates produced by our eigensolver with those obtained from the reference methods discussed above provides a relevant and practical benchmark. However, the scope of the present work extends beyond bend-loss estimation alone. The first algorithm developed below provides a mathematical and computational tool for evaluating solutions to Bessel's equation with high precision and computational efficiency in challenging parameter regimes, including large complex-valued arguments. The second and third algorithms complete the framework by enabling a reliable and accurate solution of the resulting Bessel eigenproblem, whose eigenvalues are complex-valued and large in magnitude and whose corresponding eigenfunctions are likewise complex-valued and may be highly oscillatory.}

\revC{A further major contribution of this work is the use of a perfectly matched layer (PML) to impose an improved radiation condition for the eigenproblem, leading to more reliable numerical solutions. Finally, the accurate evaluation of these eigenfunctions is essential for prescribing propagating modes as boundary data in finite element simulations of the corresponding three-dimensional electromagnetic boundary value problems and for assessing the performance of such simulations~\cite{mora2026bent}.}


\revC{The paper is organized as follows. Section~\ref{sec:background} introduces the mathematical and physical background for the circular waveguide problem. Section~\ref{sec:EvalBesselFunctions} develops algorithms based on the classical Frobenius method to evaluate the rescaled Bessel functions and their derivatives. Section~\ref{sec:3layer_circ_eig} then applies these tools to the three-layer circular waveguide with an impedance BC and presents the corresponding eigenvalue solver and numerical results. In Section~\ref{sec:PML-BC}, we replace the impedance BC with a perfectly matched layer (PML) and examine the resulting changes in the computed eigensolutions. Section~\ref{sec:BendLoss} uses the PML-based formulation to estimate bend losses and compares the results with established loss formulas from the literature. Appendix~\ref{sec:NumericalGlazmanCriterion} provides additional numerical evidence related to a hypothesis arising in the spectral analysis of non-self-adjoint operators~\cite{Demkowicz_Gopalakrishnan_Heuer_24}, thereby supporting the well-posedness considerations underlying our approach. Finally, Appendix~\ref{sec:convergence} presents further numerical results illustrating the convergence behavior of the proposed algorithms.}

%
\section{Theoretical background}
\label{sec:background}

This paper is devoted to a study of the three-layer circular waveguide problem shown in Fig.~\ref{fig:waveguide}.

\subsection{Mathematical problem definition}

\begin{figure}[htb]
    \centering
    \includegraphics[angle=0,width=.4\textwidth]{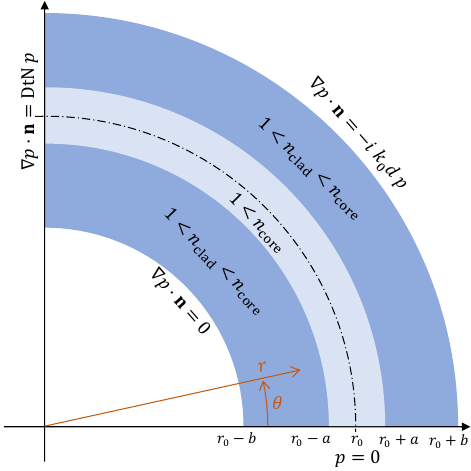}
    \caption{Geometry and boundary conditions for a circular waveguide with a core and a cladding with refractive indices $n_{\text{core}}$ and $n_{\text{clad}}$, respectively.}
    \label{fig:waveguide}
\end{figure}

We want to determine the complex-valued scalar field $p$, satisfying
\begin{itemize}
    \item Helmholtz equation in the waveguide domain $(r,\theta) \in (r_0 - b, r_0 + b) \times (0, \pi / 2)$: 
        \be
            \frac{1}{r} \frac{\ptl}{\ptl r} \left( r \frac{\ptl p}{\ptl r} \right) + \frac{1}{r^2} \frac{\ptl^2 p}{\ptl \theta^2} + k_0^2 n^2 p = 0 \, ,
            \label{eq:Helmholtz}
        \ee
        where $n = n(r)$ is the refractive index taking different values in the three layers (either $n_{\core}$ or $n_{\clad}$, see Fig.~\ref{fig:waveguide});
    \item interface conditions at $r = r_0 \pm a$:
        $$
            [\![ p ]\!] = [\![ \ptl p/\ptl r ]\!] = 0 
        $$
        with $[\![\cdot]\!]$ denoting the jump;
    \item Neumann boundary condition (BC) at $r = r_0 - b$:
        $$
            \nabla p \cdot \nml = 0 \, ,
        $$
        where $\nml$ is the outward unit normal vector;
    \item impedance (Cauchy or Robin) BC at $r = r_0 + b$:
        $$
            \nabla p \cdot \nml = - i k_0 d p \, ,
        $$
        where $i = \sqrt{-1}$, $k_0>0$ denotes the free-space wave number, and $d \geq 0$ represents the strength of the impedance condition; 
        and
    \item ``initial'' non-homogeneous Dirichlet BC at $\theta = 0$ (specifying the initial field injected into the waveguide), and a non-local Dirichlet-to-Neumann (DtN) absorbing BC at $\theta = \pi/2$. 
\end{itemize}

The derivations in this paper are made under the convention of a time-harmonic solution of the form $\exp(+i\omega t)p(r,\theta)$, with a known angular frequency $\omega>0$. Be aware that the leading sign in the exponential implies the sign of some terms of the problem's equations, particularly in the impedance BC. Now, assuming an ansatz of $p(r, \theta) = \exp (-i \beta \theta) u(r)$, where $\beta \in \bb{C}$ is known as the 
\textit{propagation constant}\footnote{We assume $\Re(\beta)>0$.} along the circumferential coordinate $\theta$, we arrive at our governing eigenproblem for the mode $u(r)$ and its eigenvalue $\lambda = \beta^2$: 
\begin{itemize}
    \item Governing mode eigenrelation: 
        \be
            r \frac{d}{dr} \left( r \frac{du}{dr} \right) + \left( \wavenum^2 r^2 - \lambda \right) u = 0 \ ,
            \label{eq:EigenProblem}
        \ee
        where the wavenumber is $\wavenum(r) \defeq k_0 n(r)$. 
        If the medium is homogeneous ($n = n_0 > 0$), then the wavenumber is also a constant, $\wavenum = k_0 n_0 > 0$, and the problem becomes a standard Bessel differential equation. 
        This equation can be obtained by rescaling the independent variable, $\tilde{r} \defeq \wavenum r$, which implies that
        \begin{align}
            \mbox{} & \dd{}{r} = \wavenum \dd{}{\tilde{r}} \ , \quad \ddn{2}{}{r} = \wavenum^{2} \frac{d^2}{d \tilde{r}^{\, 2}} \ , \quad 
            r \dd{}{r} \left( r \dd{}{r} \right) = \tilde{r} \dd{}{\tilde{r}} \left( \tilde{r} \dd{}{\tilde{r}} \right) \ ,  \text{ and so} \notag \\
            \mbox{} & \tilde{r} \dd{}{\tilde{r}} \left( \tilde{r} \dd{u}{\tilde{r}} \right) + \left( \tilde{r}^{\, 2} - \lambda \right) u = 0 \ . \label{eq:Bessel_eqn}
        \end{align}
        This relation makes it clear that any Bessel solution has an order of $\sqrt{\lambda} = \beta$ and an argument that is scaled by the wavenumber, $\wavenum$. 
    \item Interface conditions at $r = r_0 \pm a$:
        \be
            [\![u]\!] = [\![ d u/d r ]\!] = 0 \, ;
            \label{eq:IC}
        \ee
    \item Neumann boundary condition (BC) at $r = r_0 - b$ ($\nml=-\mathbf{e}_r$):
        \be
            \dd{u}{r} = 0 \, ;
            \label{eq:hardBC}
        \ee
    \item Impedance BC at $r = r_0 + b$ ($\nml=\mathbf{e}_r$):
        \be
            \dd{u}{r}  + i k_0 d u = 0 \, .
            \label{eq:impBC}
        \ee
\end{itemize}

\revC{Note that this coiled slab waveguide model does not include the \textit{photoelastic effect}~\cite{fiber1978butter, nye1985physical, rosello2016measurement}, where the bending induces a stress/strain that alters the fiber's index of refraction. Including this effect would preclude us from comparing our numerical results against analytic solutions. 
Either way, the main effect of coiling the waveguide arises from the geometry of the circular bend~\cite{marcuse1976curvature, marcuse1982influence, faustini1997bend}, and this is accounted for in our bent slab waveguide problem.}

\begin{remark}
    There are two \revC{ways} to physically interpret problem \eqref{eq:Helmholtz} and its boundary conditions. On one hand, $p$ may represent the pressure in time-harmonic acoustical waves, with a \emph{sound hard} BC on the inner boundary. Alternatively, a derivation from Maxwell's equations, under the assumptions of no free charges, no material magnetic susceptibility, and a linearly polarized electric field in a direction perpendicular to the plane $(r, \theta)$, yields an identical scalar eigenproblem, where $p$ corresponds to such a component of the time-harmonic electric field, $\mathbf{E}$. Its BC on the left-hand end is interpreted as a \emph{magnetic} BC. 
    Additionally, the impedance BC in the scalar problem can be shown to be equivalent to the impedance BC for electromagnetics. In the latter, the tangential components of $\mathbf{E}$ and the magnetic field $\mathbf{H}$ \revC{satisfy} the following relationship: 
    \begin{equation}\label{eq:EM_impBC}
        Z_{\text{clad}} \nml \times (\mathbf{H} \times \nml) + \mathbf{E} \times \nml = \mathbf{0} ,
    \end{equation}
    where $Z_{\text{clad}}$ is the impedance of the cladding (outer layer), given by $Z_{\text{clad}} = Z / n_{\text{clad}}$, where $Z$ is the free-space impedance, $Z \defeq \sqrt{\mu_0 / \epsilon_0}$, which is a function of the free-space permittivity $\epsilon_0$ and free-space permeability $\mu_0$. 
    Since $\nml=\mathbf{e}_r$ on the impedance boundary (see Fig.~\ref{fig:waveguide}), BC \eqref{eq:EM_impBC} is equivalent to \eqref{eq:impBC} because
    \begin{align*}
        \mbox{} &  \text{\revA{(i)}   the $\theta$-component of \eqref{eq:EM_impBC} implies that} & \frac{1}{n_{\text{clad}}}\sqrt{\frac{\mu_0}{\epsilon_0}}H_{\theta}+p & = 0; \\
        \mbox{} &  \text{\revA{(ii)}  by Faraday's law, } i \omega \mu_0 \mathbf{H} = - \nabla\times \mathbf{E} \text{ implies} & \frac{1}{n_{\text{clad}}}\sqrt{\frac{\mu_0}{\epsilon_0}}\frac{1}{i\omega\mu_0}\pp{p}{r}+p & = 0; \text{ and} \\
        \mbox{} &  \text{\revA{(iii)} using }\omega = k_0 / \sqrt{\epsilon_0 \mu_0} \text{ there follows} & \pp{p}{r}+ \underbrace{i k_0 n_{\text{clad}}}_{i k_0 d} p & = 0 .
    \end{align*}
\end{remark}

\subsection{Physical interpretation of the solution to the eigenproblem}

Although $\lambda$ in \eqref{eq:EigenProblem} has no direct physical meaning, after rewriting our initial ansatz for the scalar field,
{\small
\[
p(r,\theta)=\exp(-i\beta \theta)u(r)=
\underbrace{\exp(-i\Re(\beta) \theta)}_{\text{harmonic}}
\underbrace{\exp(\Im(\beta) \theta)}_{\substack{\text{decays if}\\ \Im(\beta)<0}}  u(r)=
\exp\big(-i\frac{\beta}{r_0} \underbrace{r_0\theta}_{\substack{ \text{arc length}\\\approx\,\text{longitudinal}\\ \text{ coordinate}}}\big)u(r)\,,
\]
}
we note the following:
\begin{enumerate}
    \item \revC{The radially dependent factor, $u(r)$, to which we refer below as the mode profile, gives the shape and amplitude of the full solution to \eqref{eq:Helmholtz}, and its value is modulated by that of the factor depending on $\theta$.}
    \item The propagation constant $\beta$, which is multiplied by the angle $\theta$ in radians, is nondimensional. Its real part induces a harmonic function in the circumferential direction. \revC{In contrast, the imaginary part of $\beta$, if negative, causes an exponential decay as $\theta$ grows}.
    \item More precisely, we interpret $-\Im(\beta)$ as the amplitude loss factor per radian. If $\Im(\beta)$ is very close to zero, the coiling losses are not significant, but if, for instance, $-\Im(\beta)=0.01$, a single turn ($2\pi$ radians) causes a loss of $1-\exp(-0.01\cdot 2\pi)\approx 6.1\%$, whereas 50 coils induce a loss of $1-\exp(-\pi)\approx 95.7\%$, \revC{thus resulting in the loss of} most of the signal.
    \item The imaginary part of $-\beta/r_0$ can be interpreted as an amplitude loss factor per unit length ($r_0\theta$ is the arc length at the waveguide centerline). \revC{We note that, since the curvature tends to disappear as $r_0\to\infty$, $\beta/r_0$ is expected to converge to a constant of propagation along the longitudinal coordinate of the corresponding straight waveguide.}
\end{enumerate}


%
\section{Evaluation of Bessel functions of large argument and large complex order}
\label{sec:EvalBesselFunctions}

A significant computational problem arises in the simplest version of the three-layer circular waveguide, where $n = n_{0} > 0$, because the eigenfunctions (i.e., waveguide modes) are \revC{expressed in terms of} Bessel functions whose orders are dependent on the complex-valued eigenvalues (cf.\ the Bessel eigenproblem~\eqref{eq:Bessel_eqn}). 
Furthermore, these eigenvalues scale with the bend radius ($r_{0}$) according to $\lambda \sim r_0^2$. 
Thus, for slight waveguide bends (i.e., large bend radii), the eigensolutions involve evaluating Bessel functions with both large arguments and large complex orders, for which maintaining sufficient precision is not a trivial task. 

Consider dividing equation~\eqref{eq:Bessel_eqn} by $r_0^2$: 
\begin{align}
    \frac{\tilde{r}}{r_0} \dd{}{\tilde{r}} \left( \frac{\tilde{r}}{r_0} \dd{u}{\tilde{r}} \right) + \left( \frac{\tilde{r}^2}{r_0^{2}} - \frac{\lambda}{r_0^{2}} \right) u & = 0 \notag \\
    \frac{r}{r_0} \frac{d}{dr} \left( \frac{r}{r_0} \frac{du}{dr} \right) + \left( k_0^2 n_0^2  \left( \frac{r}{r_0} \right)^2 - \frac{\lambda}{r_0^2}\right) u & = 0 \text{ for } r \in (r_0 - b, r_0 + b) \, .
    \label{eq:Bessel_eqn_divided}
\end{align}
We wish to rescale the problem in such a way that we can solve it for $\mu \defeq \lambda/r_0^2$.
This can be accomplished with a substitution of the independent variable (cf.\ conformal mapping technique~\cite{heiblum1975analysis}): 
\be
    x \defeq r_0 \ln \left( \frac{r}{r_0} \right), 
    \label{eq:MappingTox}
\ee
which implies that
\[
    \frac{r}{r_0} = \exp\left(\frac{x}{r_0}\right)
    \quad \text{ and } \quad
    \frac{r}{r_0}  \frac{d u}{dr} = \frac{r}{r_0} \frac{du}{dx} \frac{dx}{dr} = \frac{r}{r_0} \frac{du}{dx} r_0 \frac{r_0}{r} \frac{1}{r_0} = \frac{du}{dx} \, .
\]
This transforms the Bessel differential equation~\eqref{eq:Bessel_eqn_divided} into 
\be
    \frac{d^2u}{dx^2} + \left( \wavenum^2 \exp\left(\frac{2x}{r_0}\right) - \mu \right) u = 0 
    \quad \text{for } 
    x \in \left( 
    r_0 \ln \left( \frac{\revC{r_0}-b}{r_0}\right) , 
    r_0 \ln \left( \frac{\revC{r_0}+b}{r_0}\right) 
    \right) .
    \label{eq:Bessel_eqn_new}
\ee
The geometric effect induced by the circular bend is represented by the $\exp(2 x / r_{0})$ factor. 

Since the thickness of the waveguide is always much smaller than the bend radius ($b \ll r_{0}$), \revC{$r/r_0 \approx 1$}, which means that $x$ \revC{lies very close to 0}.
Using a Taylor expansion about $x = 0$, $\exp(2 x / r_{0}) \approx 1$, and so
\revC{
\[
    0 = \frac{d^2u}{dx^2} + \left( k_{0}^2 n_{0}^2 \exp\left(\frac{2x}{r_0}\right) - \mu \right) u \ \simeq\ \ddn{2}{u}{x} + \left( k_{0}^{2} n_{0}^{2} - \mu \right) u \ ,
\]
}
which is approximately the eigenproblem for a straight waveguide with $\mu$ as the mode eigenvalue. 
Indeed, in the limit as $r_{0} \to \infty$, the waveguide is no longer circularly coiled, and the eigenproblem converges to
\[
    \ddn{2}{u}{x} + \left( k_{0}^{2} n_{0}^{2} - \mu \right) u = 0 \quad \text{ for } x \in (-b, b) 
    \qquad (\text{cf.\ relation~\eqref{eq:StraightWaveguideEigenProblem}})
\]
because 
\[
    \lim_{r_{0} \to \infty} r_0 \ln\left( \frac{r_0 \pm b}{r_0} \right) = \pm b \ .
\]

Using typical parameters and design properties for optical fiber waveguides (see Section~\ref{sec:3layer_circ_eig}), if 
$r_0 = 1.3 \cdot 10^4$ and
$-0.5 \leq x \leq 0.5$ (in non-dimensional units), then
$0.999923 \leq \exp(2 x / r_{0}) \leq 1.000077$. 
Such small perturbations from unity imply that, in order to compute a proper solution of the Bessel eigenequation~\eqref{eq:Bessel_eqn_new} for large bend radii, it becomes imperative to work with high-precision floating-point arithmetic. 
By using our procedures along with ``arbitrary-precision'' floating-point arithmetic \revA{(based on the Julia implementation of the \emph{multiple-precision floating-point reliable} (MPFR) C library \cite{fousse2007mpfr}, where the user may declare floating-point type variables with a desired number of bits of working precision to enable higher precision than standard 8-byte (double) precision floating-point types)}, we are able to provide semi-analytical solutions\footnote{\revA{By semi-analytical here we mean that we are evaluating the analytical power series introduced in \eqref{eq:bessel_frobenius} up to the machine precision  of our arbitrary precision variables, so that the only approximation made in practice comes from the series truncation after guaranteeing convergence to a certain number of digits.}}
of Bessel's differential equation for the bent waveguide problems described herein.


\subsection{Computation of the rescaled Bessel functions}
The Frobenius method for evaluating a Bessel function seeks a solution of the form
\be
    u(x) = \sum_{m=0}^{\infty} c_m\, x^m \, .
    \label{eq:bessel_frobenius}
\ee
We use the Taylor expansion for the variable factor $\exp(2x/r_0)$ about $x=0$:
\[
    \exp\left( \frac{2 x}{r_0} \right) = \sum_{l = 0}^{\infty} \frac{2^l}{r_0^l \, l !} \,  x^l\, .
\]
Using these two expansions in our transformed Bessel differential equation~\eqref{eq:Bessel_eqn_new}, followed by a reorganization of the power series, we obtain the following explicit recursion relation for the coefficients, $c_m$, for the Bessel solution of order $\mu$:
\be
    (m+1)(m+2) c_{m+2} + \wavenum^2 \sum_{l=0}^m \frac{2^{m-l}}{r_0^{m-l}(m-l)!} c_l - \mu c_m = 0 \quad \text{ for } m \in \mathbb{N} \, .
    \label{eq:recursion_new}
\ee

Note that this recursion specifies the solution up to two constants: $c_0$ and $c_1$, which are both free parameters. 
\revC{Once the coefficient sequence $\{c_m\}$ is known, it is possible to compute the derivative with respect to $x$ by
$$
\dd{u}{x} = \sum_{m=1}^{\infty} m\; c_m\; x^{m-1} = c_1 + \sum_{m=0}^{\infty} (m+2)\; c_{m+2}\; x^{m+1} \, .
$$
}

\revA{The algorithm to evaluate the rescaled Bessel function's power series therefore consists of setting the first two coefficients as inputs (along with $r_0$, $\wavenum$, $\mu$), solving the recursion \eqref{eq:recursion_new}, and accumulating the sum in \eqref{eq:bessel_frobenius} until the power series converges within a given truncation tolerance $\besseltol$. This is presented in Algorithm \ref{algo:bessel_frobenius}. Notice that, at the end of each iteration, the convergence of the last two terms is verified. The rationale behind that decision is that, depending on the initial two coefficients and the bend radius, the magnitude of even and odd terms of the power series might significantly differ, so it is safer to check that both the last even and the last odd terms are sufficiently small before returning the result.  As they are quite important for future computations, this algorithm returns not only the converged power series result, therein denoted $U$, but also its derivative with respect to $x$, $U'$, and the coefficient sequence $\{c_m\}$.}

\begin{algorithm}[ht]
\revA{
\caption{Evaluation of rescaled Bessel functions through the Frobenius method}\label{algo:bessel_frobenius}
\begin{algorithmic}
\Function{BesselFrobenius}{$x,r_0,\wavenum,\mu,c_0,c_1,\besseltol$}
\State $U\gets c_0 + c_1 x$
\Comment{Initialize the power series}
\State $U'\gets c_1$
\Comment{Initialize the $x$-derivative series}
\State $m\gets 0$
\Comment{Initialize the term index}
\State $|\delta| \gets 1$
\Comment{Initialize $|\delta|$ with a number greater than $\besseltol$}
\Repeat
    \State $|\delta_{\text{pre}}| \gets |\delta|$
    \Comment{Store previous increment magnitude}
    \State $E\gets \sum_{l=0}^m \frac{2^{m-l}}{r_0^{m-l}(m-l)!} c_l$
    \Comment{Precompute the inner sum in \eqref{eq:recursion_new}}
    \State $c_{m+2} \gets \frac{\mu c_m - \wavenum^2 E}{(m+1)(m+2)}$
    \Comment{Solve the recursion \eqref{eq:recursion_new}}
    \State $\delta \gets c_{m+2}\;x^{m+2}$
    \Comment{Compute the new power-series term}
    \State $U \gets U + \delta$
    \Comment{Update the power series}
    \State $\delta' \gets (m+2)c_{m+2}\;x^{m+1}$
    \Comment{Compute the new $x$-derivative term}
    \State $U' \gets U' + \delta'$
    \Comment{Update the $x$-derivative series}
    \State $m\gets m+1$
    \Comment{Update the index for the next series term}
\Until{$|\delta_{\text{pre}}|,\,|\delta| \leq \besseltol$}
\Comment{Check convergence criterion}
\State \textbf{return} $U,\;U',\;\{c_m\}$
\EndFunction
\end{algorithmic}
}
\end{algorithm}

Let the functions $\Bfirst_{\mu}(x)$ and $\Bsecond_{\mu}(x)$ be defined by the recursion relation constants obtained from~\eqref{eq:recursion_new} when choosing $c_0 = 1$, $c_1 = 0$ and $c_0 = 0$, $c_1 = 1$, respectively. 
Specifically, the constants coincide with the value of the solution and its derivative at $x = 0$ such that the recursion generates a two-parameter family of solutions:
\be
    u(x) = C \Bfirst_{\mu}(x) + D \Bsecond_{\mu}(x) \, , \quad C,D \in \doubleIC \, .
    \label{eq:general_sol_Bessel_eqn_rescaled}
\ee

Note that our functions $\Bfirst_\mu(x(r)),\, \Bsecond_\mu(x(r))$ are two
linearly independent solutions to the Bessel equation \eqref{eq:EigenProblem}. As such, they span the same solution space as the classical pair of Bessel functions of the first and second kind 
$( J_\beta (\wavenum r),Y_\beta (\wavenum r) )$, or the pair of
Hankel functions of the first and second kind 
$( H^{(1)}_\beta (\wavenum r),H^{(2)}_\beta (\wavenum r) )$. 

The present algorithm can be used to compute Bessel functions in two different contexts:
\begin{itemize}
    \item \textbf{Evaluation of particular solutions to Bessel's equation:} if $\mu$ is known, for a given $r$ we evaluate the two fundamental solutions $\Bfirst_{\mu}(x(r))$ and\break
    $\Bsecond_{\mu}(x(r))$, truncating the series whenever the two latest increments $|c_m x^m|$ are below the desired precision. Moreover, if $C$ and $D$ are provided, we can deliver the value of the particular solution \eqref{eq:general_sol_Bessel_eqn_rescaled}.
    \item \textbf{Bessel's eigenvalue equation solver:}
    If we seek the eigenvalue $\mu$ and the eigenfunction (or mode profile), $u(x(r))$, which needs to satisfy two \revC{specified} homogeneous BCs, we use an iterative nonlinear solver to determine $C$, $D$ and $\mu$. 
    A scaling condition (such as requiring $u(x(r_0-b))=1$ or setting one of the coefficients to unity) completes the system.    
    Using the propagation constants of the corresponding straight waveguide problem \revC{as} the initial guess \revC{for} the eigenvalue is typically an effective method \revC{for converging} to the desired value of $\mu$. 
    With this setup, we only \revC{need to} evaluate $\Bfirst_{\mu}$ and $\Bsecond_{\mu}$ (and their derivatives) at the endpoints, making it an efficient and accurate method to determine the eigenvalues and modes of the problem. \revA{For the case of non-constant refractive indices, such as in our motivating problem, a few more steps are required to formulate and solve the nonlinear system. Those details are given in Section \ref{sec:3layer_circ_eig}.}
\end{itemize}

\subsection{Derivative of the rescaled Bessel function with respect to its order}
When implementing a nonlinear solver for finding the eigenvalues and modes, it is prudent to examine the derivatives of the governing eigenproblem~\eqref{eq:Bessel_eqn_new} with respect to its order, $\mu$. 
\revC{In fact, in a Newton-Raphson solver, these derivatives must enter the Jacobian matrix required to find each solution update.}

By differentiating~\eqref{eq:Bessel_eqn_new} with respect to the eigenvalue $\mu$, one obtains a non-homogeneous Bessel relation:
\be\label{eq:Bessel_der_new}
    \ddn{2}{}{x}\left[ \pp{u}{\mu} \right] + \left( \wavenum^2 \exp\left(\frac{2x}{r_0}\right) - \mu \right) \pp{u}{\mu} = u \ ,
\ee
where $u$ is still the solution of~\eqref{eq:Bessel_eqn_new}, corresponding to \revC{the} particular constants $c_0, c_1$. 
However, we will now presume that ${\p}u / {\p}\mu$ can be well represented in terms of a new series expansion,
\be
    \pp{u}{\mu} = \sum_{m=0}^\infty  b_m x^m \ .
    \label{eq:bessel_mu_frobenius}
\ee
By uniqueness of the power series, we can deduce that $b_m=\ptl c_m/\ptl\mu$. 
Note that both $b_0$ and $b_1$ vanish, which is a consequence of the fact that coefficients $c_0$ and $c_1$ are fixed independently of the value of $\mu$. 
Inserting this expansion into~\eqref{eq:Bessel_der_new} produces a recursion relation for the coefficients, $b_m$: 

\be
    b_0 = b_1 = 0; \qquad
    (m+1)(m+2) b_{m+2} + \wavenum^2 \sum_{l=0}^m \frac{2^{m-l}}{r_0^{m-l} (m-l)!} b_l - \mu b_m = c_m \quad \text{ for } m \in \mathbb{N}  \, ,
    \label{eq:recursion_b_new}
\ee
where we compute the terms of the power series until the desired precision is reached. \revA{The key difference between this recursion and \eqref{eq:recursion_new}, besides the initialization of the first two coefficients, is that the right-hand side is not zero but instead given by the coefficient $c_m$ from the series \eqref{eq:bessel_frobenius}.} 
\revC{Note that differentiating ${d}u / {d}x$ with respect to $\mu$ yields 
\begin{equation*}
    \pp{}{\mu}\left[ \dd{u}{x} \right] = 
    \dd{}{x}\left[ \pp{u}{\mu} \right] = 
    \sum_{m = 2}^{\infty} m\, b_m\, x^{m-1} = 
    \sum_{m = 0}^{\infty} (m+2)\; b_{m+2}\; x^{m+1} \, ,
\end{equation*}
which implies that knowing $\{b_m\}$ suffices to evaluate $\ptl\left[ d{u}/d{x} \right]/\ptl{\mu}$.}
\revA{Algorithm \ref{algo:bessel_frobenius_dmu} presents the pseudocode for the evaluation of the derivatives of $u$ and $d u/d x$ with respect to $\mu$, therein denoted $\partial_\mu U$ and $\partial_\mu U'$, respectively. It is important to take into account that we need here the entire coefficient sequence $\{c_m\}$ as an input, not just its first two elements.}

\begin{algorithm}[htb]
\revA{
\caption{Evaluation of rescaled Bessel function derivatives with respect to~$\mu$}
\label{algo:bessel_frobenius_dmu}
\begin{algorithmic}
\Function{BesselFrobenius\_mu}{$x,r_0,\wavenum,\mu,\{c_m\},\besseltol$}
\State $\partial_\mu U\gets 0$
\Comment{Initialize the power series}
\State $\partial_\mu U'\gets 0$
\Comment{Initialize the $x$-derivative series}
\State $m\gets 0$
\Comment{Initialize the term index}
\State $|\delta| \gets 1$
\Comment{Initialize $|\delta|$ with a number greater than $\besseltol$}
\Repeat
    \State $|\delta_{\text{pre}}| \gets |\delta|$
    \Comment{Store previous increment magnitude}
    \State $E\gets \sum_{l=0}^m \frac{2^{m-l}}{r_0^{m-l}(m-l)!} b_l$
    \Comment{Precompute the inner sum in \eqref{eq:recursion_b_new}}
    \State $b_{m+2} \gets \frac{c_m + \mu b_m - \wavenum^2 E}{(m+1)(m+2)}$
    \Comment{Solve the recursion \eqref{eq:recursion_b_new}}
    \State $\delta \gets b_{m+2}\;x^{m+2}$
    \Comment{Compute the new power-series term}
    \State $\partial_\mu U \gets \partial_\mu U + \delta$
    \Comment{Update the power series}
    \State $\delta' \gets (m+2)b_{m+2}\;x^{m+1}$
    \Comment{Compute the new $x$-derivative term}
    \State $\partial_\mu U' \gets \partial_\mu U' + \delta'$
    \Comment{Update the $x$-derivative series}
    \State $m\gets m+1$
    \Comment{Update the index for the next series term}
\Until{$|\delta_{\text{pre}}|,\,|\delta| \leq \besseltol$}
\Comment{Check convergence criterion}
\State \textbf{return} $\partial_\mu U,\;\partial_\mu U'$
\EndFunction
\end{algorithmic}
}
\end{algorithm}

\revC{If coefficients $C$ and $D$ are known, then the} partial derivative of \eqref{eq:general_sol_Bessel_eqn_rescaled} with respect to $\mu$ is given by 
\be\label{eq:der_general_sol_Bessel_eqn_rescaled}
    \pp{u}{\mu} = C \pp{\Bfirst_{\mu}}{\mu}(x)  + D \pp{\Bsecond_{\mu}}{\mu}(x) \, .
\ee
\revC{Here}, ${\p}\Bfirst_{\mu} / {\p}\mu$  is obtained by computing $b_m$ with the resulting sequence \eqref{eq:recursion_new} initialized with $c_0=1$, $c_1=0$, while $\ptl{\Bsecond_{\mu}}/\ptl{\mu}$ uses the \revC{$b_m$ and $c_m$ corresponding to} $c_0=0$, $c_1=1$.


The described algorithms are utilized to analyze the three-layer circularly bent waveguide, which must satisfy the interface conditions~\eqref{eq:IC}. 
The intricate treatment of the piecewise solution and the interface conditions is delineated below. 
\revA{Appendix~\ref{sec:NumericalGlazmanCriterion} presents an additional example, not central to this document but relevant to our research, in which we apply the present ideas to solve the homogeneous circularly coiled waveguide subject to the impedance BC \eqref{eq:impBC} on the outer boundary.}

\revC{
\begin{remark}[On the notation for derivatives]
All solutions to our eigenvalue problem and, in particular, fundamental solutions $\Bfirst_\mu(x)$ and $\Bsecond_\mu(x)$, depend upon both argument $x$ and order $\mu$. We do use the partial derivative symbol for the derivative with respect to order $\mu$, but the ordinary derivative symbol for derivatives in $x$ to explicitly relate to the ordinary differential equation form in \eqref{eq:Bessel_eqn_new}.
\end{remark}
}

%

\section{Three-layer circular waveguide with impedance BC}
\label{sec:3layer_circ_eig}

Next, we solve the three-layer waveguide eigenproblem in the $x$ coordinate. 
The solution consists of three branches corresponding to the core and the two (left and right) cladding regions. 
The solution in the waveguide core region will be represented as a function of $x$, with $x = 0$ corresponding to $r = r_0$; the solution in the left cladding region, $r < r_0 - a$, will be represented as a function of $x_1$, such that $x_1 = 0$ corresponds to $r = r_0 - a$; and the solution in the right cladding region, $r > r_0 + a$, will be represented as a function of $x_2$, such that $x_2=0$ corresponds to $r = r_0 + a$. The logarithmic transformation of $x$ is as presented in Section \ref{sec:EvalBesselFunctions}, \text{cf.~\eqref{eq:MappingTox}}, while for $x_1$ and $x_2$ we change the base point $r_0$ to $r_0-a$ and $r_0+a$, respectively:
\[
    x_{1}(r) \defeq (r_0 - a) \ln\left( \frac{r}{r_0 - a} \right) , 
    \ \
    x(r)     \defeq r_0 \ln\left( \frac{r}{r_0} \right) , 
    \ \ 
    \revC{x_{2}}(r) \defeq (r_0 + a) \ln\left( \frac{r}{r_0 + a} \right) .
\]
The regional interfaces/boundaries occur at 
\begin{align*}
    -b_1 & \defeq (r_0 - a) \ln\left( \frac{r_0 - b}{r_0 - a} \right) \, ,
    &
    -a_1 & \defeq r_0 \ln\left( \frac{r_0 - a}{r_0} \right) \ , \\
    a_2 & \defeq r_0 \ln\left( \frac{r_0 + a}{r_0} \right) \, ,
    & \text{and }
    b_2 & \defeq (r_0 +a) \ln\left( \frac{r_0 + b}{r_0 + a} \right) \, .
\end{align*}
The wavenumber $\wavenum$ for the left ($r < r_0 - a$) and right ($r > r_0 + a$) cladding regions is $\wavenum_\clad = k_0 n_\clad$, whereas for the core region it is $\wavenum_\core = k_0 n_\core$. Likewise, the eigenvalues $\mu$ become regionally defined:
\[
    \begin{cases}
            \mu_1 \defeq \lambda / (r_0 - a)^2 & \text{for } x_1(r) \in \left[ -b_1, 0 \right] , \\
            \mu \ \defeq \lambda / r_0^2 & \text{for } x(r) \in \left[ -a_1, a_2 \right] , \\
            \mu_2 \defeq \lambda / (r_0 + a)^2 & \text{for } x_2(r) \in \left[ 0, b_2 \right] .
    \end{cases}
\]

As an ansatz, the solution is presumed to take the form 
\be
    u(r)=
    \left\{
        \begin{array}{r@{}ll}
            u^{\lclad}(x_1(r)) &\ = C_1 \Bfirst^{\lclad}_{\mu_1}(x_1(r)) + D_1 \Bsecond^{\lclad}_{\mu_1}(x_1(r))   & \text{for } x_1(r) \in \left[ -b_1, 0 \right] , \\[8pt]
            u^{\core}(x(r)) &\ = C_0   \Bfirst^{\core}_{\mu}(x(r)) + D_0 \Bsecond^{\core}_{\mu}(x(r))        & \text{for } x(r) \in \left[ -a_1, a_2 \right] , \\[8pt]
            u^{\rclad}(x_2(r)) &\ = C_2 \Bfirst^{\rclad}_{\mu_2}(x_2(r)) + D_2 \Bsecond^{\rclad}_{\mu_2}(x_2(r))   & \text{for } x_2(r) \in \left[ 0, b_2 \right]\,,
        \end{array}
    \right.
    \label{eq:sol3laycirc_rescaled}
\ee
where the functions $\Bfirst_\mu$ and $\Bsecond_\mu$ 
\revC{and their derivatives with respect to $x$ (or $x_1$ or $x_2$) can be computed through Algorithm \ref{algo:bessel_frobenius}},
with the specifications detailed in Table~\ref{tab:piecewise_bessel_specs}. It is important to make sure that the substitution of the base point $r_0$ \revC{by} either $r_0-a$ or $r_0+a$ (according to the \revC{specific} cladding region), is taken into account not only when evaluating $x_1$ or $x_2$ but also within the \revC{inner sums in} recursions \eqref{eq:recursion_new} and \eqref{eq:recursion_b_new}.

\begin{table}[htb]
    \centering
    \caption{Detailed description of the argument and parameters corresponding to the evaluation of the custom fundamental Bessel solutions $\Bfirst_\mu$ and $\Bsecond_\mu$, calculated through \eqref{eq:bessel_frobenius}-\eqref{eq:recursion_new}, required to construct the piecewise solution \eqref{eq:sol3laycirc_rescaled}.}
    \label{tab:piecewise_bessel_specs}
    \begin{tabular}{|c||c|c|c|c|c|c|}
    \hline
    \multirow{2}{*}{Function $\Bfirst_\mu$ or $\Bsecond_\mu$} &
    \multicolumn{6}{c|}{Argument and parameters to evaluate} \\ \cline{2-7}
    & $x\mapsfrom$ & $r_0\mapsfrom$ & $\wavenum\mapsfrom$ & $\mu\mapsfrom$ & $c_0\mapsfrom$ & $c_1\mapsfrom$ \\ \hline
    $\Bfirst^{\lclad}_{\mu_1}$ & $x_1(r)$      & $r_0 - a$       & $\wavenum_\clad$     & $\mu_1$         & $1$             & $0$             \\ \hline
    $\Bsecond^{\lclad}_{\mu_1}$& $x_1(r)$      & $r_0 - a$       & $\wavenum_\clad$     & $\mu_1$         & $0$             & $1$             \\ \hline
    $\Bfirst^{\core}_{\mu}$  & $x(r)$        & $r_0$           & $\wavenum_\core$     & $\mu$         & $1$             & $0$             \\ \hline
    $\Bsecond^{\core}_{\mu}$ & $x(r)$        & $r_0$           & $\wavenum_\core$     & $\mu$         & $0$             & $1$             \\ \hline
    $\Bfirst^{\rclad}_{\mu_2}$ & $x_2(r)$      & $r_0 + a$       & $\wavenum_\clad$     & $\mu_2$         & $1$             & $0$             \\ \hline
    $\Bsecond^{\rclad}_{\mu_2}$& $x_2(r)$      & $r_0 + a$       & $\wavenum_\clad$     & $\mu_2$         & $0$             & $1$             \\ \hline
    \end{tabular}
\end{table}

\subsection{Treatment of interface and boundary conditions}
The interface conditions~\eqref{eq:IC} \revC{give rise to}
\be
    u^{\lclad}(0) = \revC{\underbrace{u^{\core}(-a_1)}_{C_1}} 
    \qquad\text{and}\qquad 
    u^{\rclad}(0) = \revC{\underbrace{u^{\core}(a_2)}_{C_2}} \ ,
    \label{eq:interface1}
\ee
and, noting the \revC{changes in the derivatives after switching the coordinate $r$ to $x$, $x_1$, or $x_2$},
\be
    \dd{u^{\lclad}}{x_1}(0) = 
    \revC{\underbrace{\frac{r_0}{r_0 - a}\,\dd{u^{\core}}{x}(-a_1)}_{D_1}}
    \qquad \text{and} \qquad 
    \dd{u^{\rclad}}{x_2}(0) = 
    \revC{\underbrace{\frac{r_0}{r_0 + a} \,\dd{u^{\core}}{x}(a_2)}_{D_2}}  \, .
    \label{eq:interface2}
\ee
Finally, the Neumann BC~\eqref{eq:hardBC} \revB{at the inner boundary is reformulated for the derivative of $u^{\lclad}(x_1)$ at $x_1 = -b_1$,}
\be
    \frac{r_0 - a}{r_0 - b} \, \dd{u^{\lclad}}{x_1}(-b_1) = 0 \, , \label{eq:BC1}
\ee
and the impedance BC~\eqref{eq:impBC}\revB{, now applied to $u^{\rclad}(x_2)$ at $x_2 = b_2$,} is given by 
\be
    \frac{r_0 + a}{r_0 + b} \, \dd{u^{\rclad}}{x_2}(b_2) + i \wavenum_\clad d u^{\rclad}(b_2) = 0 \, . \label{eq:BC2}
\ee

\subsection{Detailed description of the eigenvalue solver for a circularly bent three-layer waveguide}
 For our study, we seek two types of eigensolutions~(\ref{eq:sol3laycirc_rescaled}):
 \begin{itemize}
    \item \textit{Even} modes, where $C_0 \equiv 1$ and $D_0$ needs to be determined; and 
    \item \textit{Odd} modes, where $D_0 \equiv 1$ and $C_0$ needs to be determined. 
\end{itemize}
We evaluate the solution and its derivative at $x \in \{ -a_{1}, a_{2} \}$ and use interface conditions~(\ref{eq:interface1})--(\ref{eq:interface2}) to generate the corresponding solutions in the two cladding regions.  
In other words, constants $C_1,\,D_1,\,C_2,\,D_2$ are determined on-the-fly using~(\ref{eq:interface1})--(\ref{eq:interface2}), so that $u^{\lclad}$ and $u^{\rclad}$ can be constructed and evaluated. 
Finally, enforcing BCs~(\ref{eq:BC1})--(\ref{eq:BC2}) leads to a system of two nonlinear equations for unknowns $\{D_0,\lambda\}$ (when seeking even modes), or $\{C_0,\lambda\}$ (when seeking odd modes).

For example, considering an even mode ($C_0\equiv1$), we start by imposing the zero-valued Neumann BC on the inside of the bend and the impedance BC on the outside of the bend:
\[
    \left\{
    \begin{array}{lll}
        \ds \text{BC}_{\text{inner}}(D_0,\lambda) : = \dd{u^{\lclad}}{x_1} (-b_1) = 0 \ \ (\text{cf.~\ref{eq:hardBC}}) , \ \ \text{and} \\[8pt]
        \ds \text{BC}_{\text{outer}}(D_0,\lambda) := \frac{r_0+a}{r_0+b} \, \dd{u^{\rclad}}{x_2}(b_2) + i k_0 d u^{\rclad}(b_2) = 0 \ \ (\text{cf.~\ref{eq:BC2}}) .
    \end{array}
    \right.
\]
This nonlinear system ($\text{BC}_{\text{inner}}, \text{BC}_{\text{outer}}$) is solved for $D_0$ and $\lambda$ numerically using the Newton--Raphson method, requiring the derivatives of the two functions $\text{BC}_{\text{inner}}$ and $\text{BC}_{\text{outer}}$. 
Starting with $\text{BC}_{\text{inner}}$, using \revC{expressions \eqref{eq:sol3laycirc_rescaled}-\eqref{eq:interface2}}, we find that 
\begin{align*}
    \revC{\text{BC}_{\text{inner}}(D_0,\lambda)} & = C_1 \dd{\Bfirst^{\lclad}_{\mu_1}}{x_1}(-b_1) + D_1 \dd{\Bsecond^{\lclad}_{\mu_1}}{x_1}(-b_1) \\
    & = u^{\core}(-a_1) \dd{\Bfirst^{\lclad}_{\mu_1}}{x_1}(-b_1) + \frac{r_0}{r_0-a} \dd{u^{\core}}{x}(-a_1) \pp{\Bsecond^{\lclad}_{\mu_1}}{x_1}(-b_1) ,
\end{align*}
where
\[
    u^{\core}(-a_1) = C_0 \Bfirst^{\core}_{\mu}(-a_1) + D_0 \Bsecond^{\core}_{\mu}(-a_1) ,
\]
and
\[
    \dd{u^{\core}}{x}(-a_1) = C_0 \dd{\Bfirst^{\core}_{\mu}}{x}(-a_1) + D_0 \dd{\Bsecond^{\core}_{\mu}}{x}(-a_1) .
\]
Differentiating these two formulas with respect to constant $D_0$, we obtain 
\[
    \pp{}{D_0}\left[ u^{\core}(-a_1) \right] = \Bsecond^{\core}_{\mu}(-a_1) 
    \quad \text{and} \quad 
    \pp{}{D_0}\left[ \dd{u^{\core}}{x}(\revC{-a_1}) \right] = \dd{\Bsecond^{\core}_{\mu}}{x}(-a_1) .
\]
Hence,
\[
    \revC{\pp{[\text{BC}_{\text{inner}}]}{D_0}(D_0,\lambda)} = \Bsecond^{\core}_{\mu}(-a_1) \dd{\Bfirst^{\lclad}_{\mu_1}}{x_1}(-b_1) + \frac{r_0}{r_0 - a} \dd{\Bsecond^{\core}_{\mu}}{x}(-a_1) \dd{\Bsecond^{\lclad}_{\mu_1}}{x_1}(-b_1) \, .
\]
Similarly, but now differentiating with respect to the other unknown, $\lambda$, one finds that 
{\small
\begin{align*}
    \revC{\pp{[\text{BC}_{\text{inner}}]}{\lambda}} & = \pp{}{\lambda}\left[ u^{\core}(-a_1) \right] \dd{\Bfirst^{\lclad}_{\mu_1}}{x_1}(-b_1) + \frac{r_0}{r_0 - a} \pp{}{\lambda}\left[ \dd{u^{\core}}{x}(-a_1) \right] \dd{\Bsecond^{\lclad}_{\mu_1}}{x_1}(-b_1) \\
        & \hspace{13pt} + u^{\core}(-a_1) \pp{}{\lambda}\left[ \dd{\Bfirst^{\lclad}_{\mu_1}}{x_1}(-b_1) \right] + \frac{r_0}{r_0 - a} \dd{u^{\core}}{x}(-a_1) \pp{}{\lambda}\left[ \dd{\Bsecond^{\lclad}_{\mu_1}}{x_1}(-b_1) \right] \\
    & = \frac{1}{r_0^2} \left( \pp{}{\mu}\left[ u^{\core}(-a_1) \right] \dd{\Bfirst^{\lclad}_{\mu_1}}{x_1}(-b_1) + \frac{r_0}{r_0 - a} \pp{}{\mu}\left[ \dd{u^{\core}}{x}(-a_1) \right] \dd{\Bsecond^{\lclad}_{\mu_1}}{x_1}(-b_1) \right) \\
        & \hspace{13pt} + \frac{1}{(r_0 - a)^2} \Bigg( u^{\core}(-a_1) \pp{}{\mu_1}\left[ \dd{\Bfirst^{\lclad}_{\mu_1}}{x_1}(-b_1) \right] \\
        & \hspace{13pt} + \frac{r_0}{r_0 - a} \frac{d u^{\core}}{d x}(-a_1) \pp{}{\mu_1}\left[ \dd{\Bsecond^{\lclad}_{\mu_1}}{x_1}(-b_1) \right] \Bigg)
\end{align*}
}
where
\begin{align*}
    \pp{}{\mu}\left[ u^{\core}(-a_1) \right] & = C_0 \pp{\Bfirst^{\core}_{\mu}}{\mu}(-a_1) + D_0 \pp{\Bsecond^{\core}_{\mu}}{\mu}(-a_1) \quad \text{and} \\
    \pp{}{\mu}\left[ \dd{u^{\core}}{x}(-a_1) \right] & = C_0 \pp{}{\mu}\left[ \dd{\Bfirst^{\core}_{\mu}}{x}(-a_1) \right] + D_0 \pp{}{\mu}\left[ \dd{\Bsecond^{\core}_{\mu}}{x}(-a_1) \right] .
\end{align*}

\revA{We recall that Algorithm \ref{algo:bessel_frobenius}, called with the distinct parameters in Table \ref{tab:piecewise_bessel_specs}, delivers all the values of $\Bfirst_\mu$ and $\Bsecond_\mu$, along with their $x$-derivatives, that are needed to evaluate the boundary operators. Furthermore, Algorithm \ref{algo:bessel_frobenius_dmu} has been precisely tailored for the evaluation of all the derivatives with respect to $\mu$ and $\mu_1$ appearing in the above computation of $\partial[\text{BC}_{\text{inner}}]/\partial \lambda$.}
 
Analogously, one determines the formulas for the derivatives of $\text{BC}_{\text{outer}}$ with respect to the unknowns $\lambda$ \revC{(which will in practice require derivatives with respect to $\mu$ and $\mu_2$) and $D_0$, so that we can use them in Newton's method.} \revA{We show the pseudocode for the eigensolver applied to an even mode in Algorithm \ref{algo:eigensolver-3layer}.}
In the case of an odd mode, we require the partial derivatives $\ptl/\ptl C_0$ instead of $\ptl/\ptl D_0$, because in that case $C_0$ is the unknown while $D_0 = 1$ is fixed.
\begin{algorithm}[htb]
\revA{
\caption{Eigensolver for an even mode of the three-layer circular waveguide}\label{algo:eigensolver-3layer}
\begin{algorithmic}
\setlength{\itemsep}{0.9ex}
\Function{SolveEvenMode}{$r_0,a,b,\wavenum_\core,\wavenum_\clad,D_{0}^\text{guess},\lambda^{\text{guess}},\eigentol,\besseltol$}
\State $M_{\text{ref}} \gets \|(D_{0}^\text{guess},\lambda^{\text{guess}})\|$
\Comment{Store norm of initial guess as a reference}
\State $(D_{0},\lambda) \gets\ (D_{0}^\text{guess},\lambda^{\text{guess}})$
\Comment{Initialize solution with initial guess}
\Repeat
\Comment{Solve nonlinear system with Newton--Raphson}
\State Evaluate $\, \text{BC}_{\text{inner}}\, $, $\ \partial[\text{BC}_{\text{inner}}]/\partial D_0\, $, and $\, \partial[\text{BC}_{\text{inner}}]/\partial \lambda\ $ at current $(D_0,\lambda)$.
\State Evaluate $\, \text{BC}_{\text{outer}}\, $, $\ \partial[\text{BC}_{\text{outer}}]/\partial D_0\, $, and $\, \partial[\text{BC}_{\text{outer}}]/\partial \lambda\ $ at current $(D_0,\lambda)$.
\State Assemble Jacobian matrix 
$\bfJ = 
\begin{pmatrix}
    \partial[\text{BC}_{\text{inner}}]/\partial D_0 & \partial[\text{BC}_{\text{inner}}]/\partial \lambda \\
    \partial[\text{BC}_{\text{outer}}]/\partial D_0 & \partial[\text{BC}_{\text{outer}}]/\partial \lambda \\
\end{pmatrix}$.
\State Assemble residual vector 
$\bfR=(-\text{BC}_{\text{inner}}\;,\ -\text{BC}_{\text{outer}})^T$.
\State Solve for the increment $\Delta\bfS$ in the system $\bfJ \Delta\bfS = \bfR$.
\State $(D_0\,,\, \lambda)^T \gets (D_0\,,\, \lambda)^T + \Delta\bfS$.
\Comment{Update the solution}
\Until{ $ \| \Delta \bfS \| \leq \eigentol  M_{\text{ref}}$\ \ or the maximum number of iterations is reached.}
\State \textbf{return} $(D_0\,,\, \lambda)$
\EndFunction
\end{algorithmic}
}
\end{algorithm}
\subsection{Nondimensionalization}
The physical scales involved in the three-layer optical waveguide model, when expressed in SI units~(see Table~\ref{tab:data}), range over 26 orders of magnitude.
With sagacious choices for the characteristic sizes of the length, mass, time, and electrical current parameters, we can reduce this range to 6 orders of magnitude. 
The practical length unit is taken to match the wave\-guide's core diameter of 25.4~microns ($10^{-6}$~m), while the rest are chosen to get the magnitude of most parameters closer to 1, as shown in Table \ref{tab:nondimensional}. 
By normalizing with these four fundamental quantities, we can nondimensionalize the entire problem since the other derived units involved, such as \revC{newton (N), volt (V), and farad (F)}, can be expressed in terms of the four units given (m, kg, s, A).
\begin{table}[htb]
    \centering
    \caption{Characteristic sizes used for the nondimensionalization of the governing system.
    }
    \label{tab:nondimensional}
    \begin{tabular}
        {|l|r|}\hline
        Reference length              & $25.4 \cdot 10^{-6}$ m \\
        Reference mass                & $10^{-22}$ kg \\
        Reference time                & $10^{-13}$ s \\
        Reference electrical current  & $400$ A \\
        \hline
    \end{tabular}
\end{table}
\begin{table}[htb]
    \centering
    \caption{Coiled waveguide parameters in both SI and nondimensional units. All the waveguide's properties are adapted from the optical step-index fiber analyzed in \cite{henneking2021fiber}.}
    \label{tab:data}
    \begin{tabular}{|r|c|c|c|}
        \hline
        Parameter                       & Symbol       & Value in SI Units              & Nondimensional Value \\
        \hline
        \scalebox{0.75}{Bend radius range} & $r_0$      & \scalebox{0.75}{$[0.03302\text{ m}, 0.3302\text{ m}]$}          & \scalebox{0.75}{$[1300, 13000]$}  \\
        \scalebox{0.75}{Core's half-width}              & $a$            & \scalebox{0.75}{$1.27 \cdot 10^{-5}$ m }       & \scalebox{0.75}{$0.5$} \\
        \scalebox{0.75}{Cladding's half-width}          & $b$            & \scalebox{0.75}{$1.27 \cdot 10^{-4}$ m }       & \scalebox{0.75}{$5.0$} \\
        \scalebox{0.75}{Core's refractive index}        & $n_0$        & \scalebox{0.75}{1.4512}                          & \scalebox{0.75}{$1.4512$}\\
        \scalebox{0.75}{Cladding's refractive index}    & $n_1$        & \scalebox{0.75}{1.45}                            & \scalebox{0.75}{$1.45$}\\
        \scalebox{0.75}{Impedance BC strength}           & $d$          & \scalebox{0.75}{$1.45$}                         & \scalebox{0.75}{$1.45$}  \\
        \scalebox{0.75}{Free-space wavenumber}        & $k_0$        & \scalebox{0.75}{$5.905 \cdot 10^{6}$ rad/m }       & \scalebox{0.75}{$1.49993333460866 \cdot 10^{2}$} \\
        \scalebox{0.75}{Angular frequency}           & $\omega$     & \scalebox{0.75}{$1.77035 \cdot 10^{15}$ rad/s}      & \scalebox{0.75}{$1.77034921739554 \cdot 10^{2}$} \\
        \scalebox{0.75}{Vacuum permeability}            & $\mu_0$      & \scalebox{0.75}{$1.25664 \cdot 10^{-6}$ N/A$^2$} & \scalebox{0.75}{$0.791582400800000$} \\
        \scalebox{0.75}{Vacuum permittivity}            & $\epsilon_0$ & \scalebox{0.75}{$8.85419 \cdot 10^{-12}$ F/m}    & \scalebox{0.75}{$0.906838390340768$} \\
        \scalebox{0.75}{Free-space impedance}  & $Z$          & \scalebox{0.75}{$3.76730313412 \cdot 10^{2}$ V/A }        & \scalebox{0.75}{$0.934293045847926$}\\
        \hline
    \end{tabular}
\end{table}

From this point onward, all parameter values will be presented in their nondimensional form without notationally distinguishing them from their dimensional counterparts.

\subsection{On the computational implementation and precision}
Since we are aware that the expected eigenvalues demand outstanding accuracy, the methodology explained above has been translated \revC{into} high-precision arithmetic scientific computing codes in both Fortran and the Julia language \cite{Julia-2017}.

In Fortran, we have used 128-bit floating-point (quadruple precision) variables. With those variables, it is possible to handle floating-point arithmetic with up to thirty-four digits of precision, so we choose to enforce a precision of $10^{-32}$ in the Bessel-Frobenius power series convergence, and at most $10^{-29}$ for the eigenvalue convergence. Only the results presented in Appendix~\ref{sec:NumericalGlazmanCriterion} and in Table \ref{tab:straight_eigenvalues} are calculated with our Fortran implementation.

In the Julia implementation, thanks to the \texttt{BigFloat} type, based on the MPFR library originally developed in C, arbitrary floating-point precision is attainable. The results provided in this paper are computed with a 70-decimal-digit setup, with a Bessel power series tolerance of $10^{-65}$ and the eigenvalue tolerance set to $10^{-60}$. The maximum number of terms in the power series is set to 1000, and the maximum number of Newton iterations is 50. Extending to a more demanding precision level is straightforward, without a dramatic performance \revC{penalty}. \revC{A repository with this implementation is available as open-source software (\url{https://github.com/cgt3/WaveguideSolver}) for purposes of reproducibility of our results. Additional information on our computational setup is provided below.}

\revA{We characterize a computation as successful when all power-series evaluations and the nonlinear eigensolver (Algorithms \ref{algo:bessel_frobenius}, \ref{algo:bessel_frobenius_dmu}, and \ref{algo:eigensolver-3layer}) satisfy their prescribed convergence criteria without exceeding the specified maximum numbers of series terms or nonlinear iterations, and without exhausting the available computational resources. All numerical results reported below were obtained from computations satisfying these criteria. Conversely, we regard a computation as unsuccessful if any of the algorithms fails to converge within the prescribed limits. During the development of the implementation, we encountered such failures when the working precision was insufficient to accurately resolve the intermediate quantities entering the power-series evaluations and the nonlinear solve. Increasing the floating-point precision, together with tightening the associated convergence tolerances, eliminated these failures. Based on these tests, we selected 70 digits of working precision, a power-series tolerance of $10^{-65}$, and an eigenvalue tolerance of $10^{-60}$ for the computations reported in this paper. With these settings, all reported cases converged consistently. This behavior provides practical numerical evidence for the need to employ arbitrary-precision arithmetic in the present problem.}

\revC{Appendix \ref{sec:convergence} further illustrates the convergence behavior of the rescaled power-series evaluations and the nonlinear eigensolver under these choices of working precision and convergence tolerances.}

\subsection{How to interpret the numerical results}
In the tables of this section and the next section we report the eigenvalue $\lambda$, the propagation constant $\beta=\sqrt{\lambda}$ and the result $\beta/r_0=\sqrt{\mu}$. 
\revA{We must emphasize the point that we do not have an exact (analytical) solution or highly accurate (reliable) numerical solutions from other sources/methods for the expected eigenvalues $\lambda$ (or $\mu$, or $\beta$) of the circular waveguide problem. 
However, because these eigenvalues can be interpreted as perturbations of the straight-waveguide eigenvalues, we use the latter as reference values (see Table~\ref{tab:straight_eigenvalues}).} 
For all three modes, the reader may verify the similarity between $\sqrt{\mu}=\beta/r_0$ in Tables~\ref{tab:eig13000}, \ref{tab:eigval3_vs_r0-imp}, \ref{tab:eigval3_vs_r0-pml} and the values of $\sqrt{\mustraight}$ in Table~\ref{tab:straight_eigenvalues}, as well as the rapid growth of $-\Im(\beta)$ as the bending becomes tighter (that is, as $r_0$ decreases).

\begin{remark}
    Since one of the goals of the present research effort is to obtain benchmarking solutions for a finite element software implemented in double precision, the main results in the tables below are delivered in a fifteen-digit format, so that they can be more directly translated into a double-precision implementation of \revC{Algorithm \ref{algo:bessel_frobenius}, invoked with one of the eigenvalues computed with the arbitrary precision implementation}.
\end{remark}

\begin{remark}
    As explained along the algorithm derivations above, in addition to the eigenvalue, the unspecified coefficient $D_0$ (or $C_0$ for odd modes) is an unknown that must also be solved for. Of course, in the process of computing the numerical experiments herein published, we do solve for both unknowns and get a converged solution for such a coefficient too. We choose not to report its value in the tables below, to focus more on the eigenvalue results, as the latter are of higher interest for the physical application under investigation. However, in the accompanying code, users will get the values of $C_0$ and $D_0$ as additional outputs from the scripts that reproduce the present results.
\end{remark}

\subsection{Numerical results using the impedance BC} 
In all of our numerical experiments, we use the waveguide data provided in Table~\ref{tab:data}. 
Using the solution~\eqref{eq:sol3laycirc_rescaled} to the eigenproblem~\eqref{eq:Bessel_eqn_new}, and its accompanying numerical method for finding the unknown coefficients, we compute the eigenvalues $\lambda$ for the only three propagating modes: $\{$first even mode, first odd mode, second even mode$\}$. 
In order to determine an appropriate initial guess for $\lambda$ in the coiled waveguide problem, we first solve the straight waveguide problem:\footnote{Finding the eigenvalues of this straight waveguide problem requires solving transcendental equations that involve hyperbolic and trigonometric functions, which are derived so that all the interface and boundary conditions are satisfied~\cite{Demkowicz_Gopalakrishnan_Heuer_24}.}
\begin{equation}\label{eq:StraightWaveguideEigenProblem}
    \begin{split}
    u'' + (k_0^2 n^2 - \mustraight)u = 0 ,\\ 
    [\![u(\pm a)]\!] = 0, \ \ [\![u'(\pm a)]\!] = 0 , \\ 
    u'(-b) = 0, \ \ u'(b) - i d k_0 n(b) u(b) = 0 .
    \end{split}
\end{equation}
For both cases, $d = 0$ and $d = 1.45$, Table~\ref{tab:straight_eigenvalues} contains the predicted values of $\mustraight$ and its square root, which will be compared with $\beta / r_0$ in our forthcoming results for the circularly coiled waveguides. 
Notice that the difference between the results of these two cases is restricted to the imaginary part, and is many orders of magnitude smaller than their real parts. 

We moreover mention that these results for the straight waveguide have been obtained from a quadruple-precision implementation, making any result whose real and imaginary parts differ by more than 30 orders of magnitude \revA{(which is more common as $r_0$ grows)} unreliable. However, it was observed that despite the inaccuracy of the \revA{straight case's} imaginary part, in \revA{our} Fortran implementation \revA{the circular case's convergence benefits} from that non-real starting point, while the Neumann BC guess may cause a failed convergence. The Julia code \revA{(with 70-digit precision)}, in turn, can \revC{use either initial guess} ($d=0$ or $d=1.45$) to converge to the same result in a similar number of Newton iterations.

\begin{table}[htb]
    \centering
    \caption{Computed eigenvalues for three propagating modes in a \textbf{straight} three-layer waveguide using Neumann ($d = 0$) or impedance ($d = 1.45$) BCs on the right boundary. The digits, or entire values, of the impedance BC's results that differ from the Neumann BC's are shaded. \revA{These eigenvalues are used as initial guesses for the bent waveguide case, and serve as reference values.}
    }
    \label{tab:straight_eigenvalues}
    {\small
    \begin{tabular}{|l|l|l|l|l|l|}
        \hline
        \multicolumn{2}{|c|}{}        & \multicolumn{2}{c|}{$d=0$}               & \multicolumn{2}{c|}{$d=1.45$}            \\
        \hline
        Mode            &         & $\mustraight$ & $\sqrt{\mustraight}$ & $\mustraight$ & $\sqrt{\mustraight}$ \\
        \hline
        $1^{\text{st}}$ & ${\Re}$ & \scalebox{0.63}{$4.73785763924115 \cdot 10^{4}$} & \scalebox{0.63}{$2.17666204065793\cdot 10^{2}$} & \scalebox{0.63}{$4.73785763924115\cdot 10^{4}$}  & \scalebox{0.63}{$2.17666204065793\cdot 10^{2}$} \\
        Even            & ${\Im}$ & \scalebox{0.63}{{$0$}} & \scalebox{0.63}{{$0$}}                               & \scalebox{0.63}{\hlgray{$-1.28929634312106\cdot 10^{-34}$}} & \scalebox{0.63}{\hlgray{$-2.9616364852197\cdot 10^{-37}$}} \\
        \hline
        $1^{\text{st}}$ & ${\Re}$ & \scalebox{0.63}{$4.73594553855486 \cdot 10^{4}$} & \scalebox{0.63}{$2.17622276859582\cdot 10^{2}$} & \scalebox{0.63}{$4.73594553855486\cdot 10^{4}$} & \scalebox{0.63}{$2.176\hlgray{22276859582}\cdot 10^{2}$} \\
        Odd             & ${\Im}$ & \scalebox{0.63}{{$0$}} & \scalebox{0.63}{{$0$}}                               & \scalebox{0.63}{\hlgray{$-2.12986190583761\cdot 10^{-29}$}} & \scalebox{0.63}{\hlgray{$-4.89348318695304\cdot 10^{-32}$}} \\
        \hline
        $2^{\text{nd}}$ & ${\Re}$ &\scalebox{0.63}{$4.73251454095355 \cdot 10^{4}$} & \scalebox{0.63}{$2.17543433386383\cdot 10^{2}$}  & \scalebox{0.63}{$4.73251454095355\cdot 10^{4}$}& \scalebox{0.63}{$2.17543433386383\cdot 10^{2}$} \\
        Even            & ${\Im}$ & \scalebox{0.63}{{$0$}} & \scalebox{0.63}{{$0$}}                               & \scalebox{0.63}{\hlgray{$-6.92081537949382\cdot 10^{-20}$}} & \scalebox{0.63}{\hlgray{$-1.59067439355929\cdot 10^{-22}$}} \\
        \hline
    \end{tabular}
}
\end{table}

\begin{figure}[htb]
    \centering
    \begin{subfigure}{0.33\textwidth}
        \includegraphics[width=\textwidth]{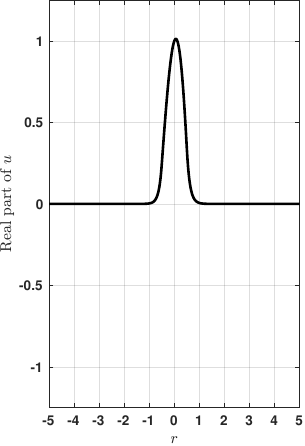}
        \caption{First even mode profile}
        \label{plot:1stEvenMode-ImpBC-r0=13000}
    \end{subfigure}%
    \begin{subfigure}{0.33\textwidth}
        \includegraphics[width=\textwidth]{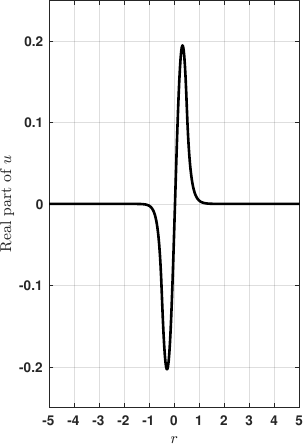}
        \caption{First odd mode profile}
        \label{plot:1stOddMode-ImpBC-r0=13000}
    \end{subfigure}%
    \begin{subfigure}{0.33\textwidth}
        \includegraphics[width=\textwidth]{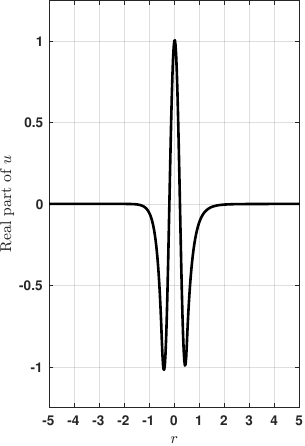}
        \caption{Second even mode profile}
        \label{plot:2ndEvenMode-ImpBC-r0=13000}
    \end{subfigure}
    \caption{Real components of the mode profiles of the three-layer waveguide~\eqref{eq:Bessel_eqn_new} at $r_0 = 13000$ with impedance BC~\eqref{eq:BC2}, which are nearly indistinguishable from \revC{those of} the straight waveguide case. Notice that the coordinate $r$ has been re-centered at zero.
    See Table~\ref{tab:eig13000} for their corresponding eigenvalues.}
    \label{fig:almost-straight-modes}
\end{figure}

\revC{To initialize the coiled waveguide eigenproblem solver with a given $r_0$, the eigenvalues $\mustraight$ of Table \ref{tab:straight_eigenvalues} are multiplied by $1.0001 \cdot r_0^2$. The values of the $d = 1.45$ case are used as seeds for the present set of results (i.e., with impedance BC), while those of $d = 0$ are utilized for the results in the next two sections (i.e., without impedance BC).}
Using our largest bend radius, $r_0 = 13000$, the resulting mode profiles are depicted in Fig.~\ref{fig:almost-straight-modes} and their corresponding eigenvalues are delineated in Table \ref{tab:eig13000}.
Because of this very weak bending, the mode profiles appear nearly identical to those of a straight waveguide \revC{and are} highly symmetric/antisymmetric about the center of the waveguide. 
However, their eigenvalues do exhibit some minuscule losses (${\Im}(\beta) < 0$) induced by this slight circular coiling. According to the above suggested interpretation of the numerical results for $\beta$, at this bending curvature it may take millions of coils to get a perceivable loss in the second even mode and many more orders of magnitude to affect the amplitude of the first even and odd modes.

\begin{table}[htb]
    \centering
    \caption{\revA{Eigenvalues} of the first two even modes and the first odd mode of~\eqref{eq:Bessel_eqn_new} at $r_0 = 13000$ using the impedance BC~\eqref{eq:BC2}. See Fig.~\ref{fig:almost-straight-modes} for their corresponding mode profiles. \revA{Compare} the last column to the straight waveguide's eigenvalues (see~Table~\ref{tab:straight_eigenvalues}) \revA{and} note that the leading figures of the real part remain equal, but the imaginary parts are some orders of magnitude larger (the digits that differ are shaded).}
    \label{tab:eig13000}
    {\small
    \begin{tabular}{|p{1.6cm}|p{0.38cm}|p{2.95cm}|p{2.95cm}|p{2.95cm}|}
        \hline
        $r_0 = 13000$ & \mbox{} & $\lambda$ & $\beta = \sqrt{\lambda}$ & $\beta/r_0=\sqrt{\mu}$ \rule[-1.5mm]{0mm}{6mm} \\
        \hline
        \multirow{2}{1.6cm}{$1^{\text{st}}$ even mode} & ${\Re}$ & \scalebox{0.75}{$8.00620263404956 \cdot 10^{12}$} 
        & \scalebox{0.75}{$2.82952339344448 \cdot 10^{6}$} & \scalebox{0.75}{$2.176\hlgray{55645649575} \cdot 10^{2}$} \\
        \mbox{} & ${\Im}$ & \scalebox{0.75}{$-2.62359245486257 \cdot 10^{-21}$} & \scalebox{0.75}{$-4.63610313479115 \cdot 10^{-28}$} & \scalebox{0.75}{\hlgray{$-3.56623318060858 \cdot 10^{-32}$}} \\
        \hline
        \multirow{2}{1.6cm}{$1^{\text{st}}$ odd mode} & ${\Re}$ & \scalebox{0.75}{$8.00294378462047 \cdot 10^{12}$} & \scalebox{0.75}{$2.82894746939926 \cdot 10^{6}$} & \scalebox{0.75}{$2.176\hlgray{11343799943}\cdot 10^{2}$} \\
        \mbox{} & ${\Im}$ & \scalebox{0.75}{$-4.15317011697652 \cdot 10^{-15}$} & \scalebox{0.75}{$-7.34048645636122 \cdot 10^{-22}$} & \scalebox{0.75}{\hlgray{$-5.64652804335478 \cdot 10^{-26}$}} \\
        \hline
        \multirow{2}{1.6cm}{$2^{\text{nd}}$ even mode} & ${\Re}$ & \scalebox{0.75}{$7.99795845391453 \cdot 10^{12}$} & \scalebox{0.75}{$2.82806620394830 \cdot 10^{6}$} & \scalebox{0.75}{$2.17543\hlgray{554149869}\cdot 10^{2}$} \\
        \mbox{} & ${\Im}$ & \scalebox{0.75}{$-2.21912100071211$} & \scalebox{0.75}{$-3.92338941290335 \cdot 10^{-7}$} & \scalebox{0.75}{\hlgray{$-3.01799185607950 \cdot 10^{-11}$}} \\
        \hline
    \end{tabular}
    }
\end{table}

\begin{figure}[htb]
    \centering
    \begin{subfigure}{0.33\textwidth}
        \includegraphics[width=\linewidth]{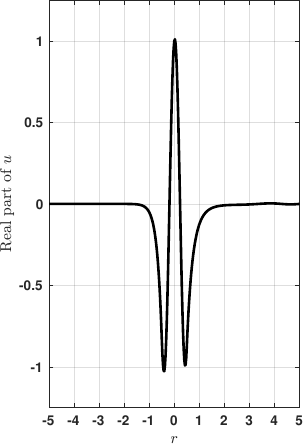}
        \caption{$r_0 = 10400$}
        \label{plot:2ndEvenMode-ImpBC-r0=10400}
    \end{subfigure}%
    \begin{subfigure}{0.33\textwidth}
        \includegraphics[width=\linewidth]{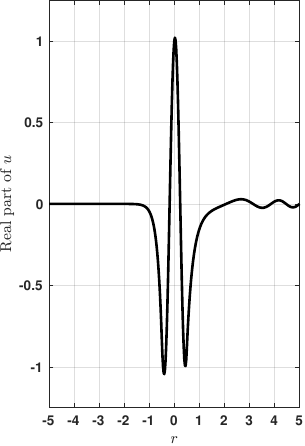}
        \caption{$r_0 = 7800$}
        \label{plot:2ndEvenMode-ImpBC-r0=7800}
    \end{subfigure}%
    \begin{subfigure}{0.33\textwidth}
        \includegraphics[width=\linewidth]{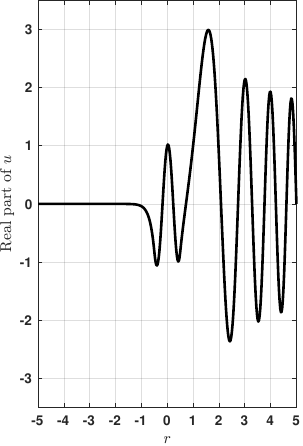}
        \caption{$r_0 = 5200$}
        \label{plot:2ndEvenMode-ImpBC-r0=5200}
    \end{subfigure}
    \caption{Real component of the second even mode's profile in the three-layer bent waveguide when using the impedance BC~\eqref{eq:BC2}, for three distinct bend radii $r_0$. Examine the data in Table~\ref{tab:eigval3_vs_r0-imp} for their corresponding eigenvalue information. Compare these mode profiles with those portrayed in Fig.~\ref{fig:mode3_vs_r0-pml}, which are computed using a PML BC.}
    \label{fig:mode3_vs_r0-imp}
\end{figure}

Focusing on the second even mode, which is more susceptible to the coiling effect, and tightening the bend radius ($r_0$), one first sees some oscillations on the outside of the bend, which grow in magnitude as the coil radius is decreased (see Fig.~\ref{fig:mode3_vs_r0-imp}). 
The results listed in Table \ref{tab:eigval3_vs_r0-imp} show that the predicted eigenvalues are becoming more lossy as the bend radius decreases; this result is expected since the bending reduces the waveguide's ability to entirely guide the modes. 
Rather, some mode energy leaks out and the mode develops a confinement loss. 
Compare these results with those of the model's predicted solutions when using a PML BC, found in Fig.~\ref{fig:mode3_vs_r0-pml} and Table~\ref{tab:eigval3_vs_r0-pml}. 

\begin{table}[htb]
    \centering
    \caption{Model-estimated eigenvalues, using the impedance BC~\eqref{eq:BC2}, of the second even mode as the coiling radius is tightened. Figure~\ref{fig:mode3_vs_r0-imp} illustrates their corresponding mode profiles. Compare these model-predicted mode eigenvalues with those delineated in Table~\ref{tab:eigval3_vs_r0-pml}, which were computed using a PML BC; the digits, or entire values, that differ are shaded.}
    \label{tab:eigval3_vs_r0-imp}
    {\small
    \begin{tabular}{|p{1.6cm}|p{0.38cm}|p{2.95cm}|p{2.95cm}|p{2.95cm}|}
        \hline
        $r_0$ & \mbox{} & $\lambda$ & $\beta = \sqrt{\lambda}$  & $\beta/r_0=\sqrt{\mu}$ \rule[-1.5mm]{0mm}{6mm} \\
        \hline
        \multirow{2}{1.6cm}{$10400$} & ${\Re}$ & \scalebox{0.75}{$5.11869686\hlgray{714306} \cdot 10^{12}$} & 
        \scalebox{0.75}{$2.26245372\hlgray{707224} \cdot 10^{6}$} & \scalebox{0.75}{$2.17543627\hlgray{603100}\cdot 10^{2}$} \\
        \mbox{} & ${\Im}$ & \scalebox{0.75}{\hlgray{$-1.21328979840387 \cdot 10^{2}$}} & \scalebox{0.75}{\hlgray{$-2.68135826135536 \cdot 10^{-5}$}} & \scalebox{0.75}{$\hlgray{-2.57822909745708} \cdot 10^{-9}$} \\
        \hline
        \multirow{2}{1.6cm}{$7800$} & ${\Re}$ & \scalebox{0.75}{$2.879271\hlgray{57786418} \cdot 10^{12}$} & 
        \scalebox{0.75}{$1.6968416478\hlgray{4584} \cdot 10^{6}$} & \scalebox{0.75}{$2.1754380\hlgray{1005877}\cdot 10^{2}$} \\
        \mbox{} & ${\Im}$ & \scalebox{0.75}{\hlgray{$-5.76638397398440 \cdot 10^{3}$}} & \scalebox{0.75}{\hlgray{$-1.69915206327735 \cdot 10^{-3}$}} & \scalebox{0.75}{\hlgray{$-2.17840008112481 \cdot 10^{-7}$}} \\
        \hline
        \multirow{2}{1.6cm}{$5200$} & ${\Re}$ & \scalebox{0.75}{$1.2796\hlgray{2392974873} \cdot 10^{12}$} & 
        \scalebox{0.75}{$1.1312\hlgray{0463654890} \cdot 10^{6}$} & \scalebox{0.75}{$2.175\hlgray{39353182480}\cdot 10^{2}$} \\
        \mbox{} & ${\Im}$ & \scalebox{0.75}{$-\hlgray{2.25075256570834} \cdot 10^{6}$} & \scalebox{0.75}{$-0.\hlgray{994847657526836}$} & \scalebox{0.75}{$-1.\hlgray{91316857216699} \cdot 10^{-4}$} \\
        \hline
    \end{tabular}
    }
\end{table}


%
\section{Three-layer circular waveguide with a PML}
\label{sec:PML-BC}

It is well known that the impedance BC is a good approximation of the radiation condition provided that the outgoing wave is exiting mostly in the normal direction to the surface of the boundary. 
For the three-layer circularly coiled waveguide problem, the wave predominantly exits in the tangential direction,
so the impedance BC's validity for approximating a proper radiation condition (as in an open waveguide) is at best questionable.
Presently, we will demonstrate that the impedance BC can be replaced with a much more accurate \emph{perfectly matched layer} (PML).

\begin{figure}[htb]
    \centering
    \includegraphics[width=0.6\textwidth]{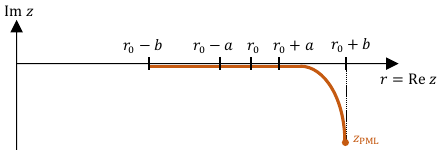}
    \caption{Passing from a segment in the real line to a path in the complex plane, inspired by the PML methodology. The goal of extending the solution to the complex plane is to capture the decay that the outgoing wave function is expected to present when evaluated on the (orange) curved path. Such decaying behavior is enforced through the constraint $u=0$ at $z=\zpml$.
    }
    \label{fig:PML}
\end{figure}

The construction of a PML BC can be accomplished in a few steps and is supported by the following arguments:
\begin{itemize}
    \item The solution is holomorphic in $r$ and thus admits a holomorphic extension. In simple \revC{terms}, it is allowed to replace the independent real-valued variable $r$ with a complex-valued argument $z$, so that the derivatives in the original Bessel equation~\eqref{eq:Bessel_eqn} are then understood in the complex sense. 
    \item  The original solution has been traced along $r = \Re(z)$. This makes differentiating between the outgoing and incoming waves impossible, until the imaginary component of $z$ is reintroduced. Therefore, by having $z$ dive into the complex plane in a layer near the boundary (PML), as is depicted in Fig.~\ref{fig:PML}, we are able to distinguish ingoing and outgoing waves, thanks to the mathematical argumentation outlined next.
    \item Marcuse~\cite[pp. 408]{marcuse1982light} shows that the outgoing solution in the outer cladding is a Hankel function of the second kind, of order $\beta$, with an argument scaled by $k_0 n_{\text{clad}}$, i.e., $H^{(2)}_\beta(k_0 n_{\text{clad}} z)$. 
    According to the asymptotic properties of these Hankel functions (see \cite[\S10.2,\S10.7]{NIST:DLMF} or \cite[\S8.451]{gradshteyn2014table}), we have
    \be
        H^{(2)}_\beta(k_0 n_{\text{clad}}z)\approx \sqrt{\frac{2}{\pi k_0 n_{\text{clad}}z}}\exp\{-i(k_0 n_{\text{clad}}z-\beta\pi/2-\pi/4)\}
        \label{eq:hankel_asymptotics}
    \ee
    with the truncation error, in \emph{little-o} notation, being estimated by \break $o(|\Im(k_0 n_{\text{clad}}z)|)$ as $z \to \infty$. 
    The asymptotic expression for the corresponding Hankel function of the first kind, $H^{(1)}_\beta$, which models an incoming wave at the boundary, differs only by the leading sign in the exponential. 
    By tracing the solution of our Bessel equation along a path where $z$ has an increasingly negative imaginary part, outgoing waves (i.e., $H^{(2)}_\beta$ functions) decay exponentially, whereas incoming waves (i.e., $H^{(1)}_\beta$ functions) grow exponentially. 
    \item In order to implement the PML in practice, we designate the new path in the boundary layer by the formula $z = z(r)$, and appropriately modify the governing relation~\eqref{eq:Bessel_eqn}. It is important to realize that the projection of the solution onto power series expansions remains valid even after implementing the complex argument $z$ in the boundary layer  (see Remark~\ref{remark:OrderOfTransformations} about the additional coordinate transformation applied to $x$).
    The new complex-valued path $z(r)$ needs to start at a point on the real line between $r_0+a$ and $r_0+b$ and to converge smoothly to a preset point $\zpml$ (see Fig.~\ref{fig:PML}), which has a negative imaginary part that depends on the cladding wavenumber $\wavenum_\clad$ and a positive real parameter $\pmlparameter$, which controls the strength of the PML absorption:
    \begin{equation}
        \zpml = r_0 + b - i\, \frac{\pmlparameter}{\wavenum_\clad}\, ,
        \label{eq:zpml_definition}
    \end{equation}
    In the PML literature, the path toward the point \eqref{eq:zpml_definition} is usually a polynomial curve.
    At the end of this layer (at $r = r_0 + b$), the impedance BC is replaced with a zero-valued Dirichlet BC for the modified governing eigenequation (in terms of $z$):
    \begin{equation}
    u(\zpml) = 0 .
    \label{eq:PML_bc}
    \end{equation}
    The Dirichlet BC enforced at $z = \zpml$ eliminates incoming waves while leaving outgoing waves unaffected (up to machine precision), because, at that point, the latter are mostly attenuated/absorbed. Moreover, the choice of a particular path through the PML does not matter because
    the Dirichlet BC is enforced only at the final point. \revC{The algorithm} never evaluates the power series at other points along the path and makes no use of the particular curve parameterization (e.g., derivatives with respect to $z$ are not required for computations). Finally, notice that replacing the impedance BC with \eqref{eq:PML_bc} significantly simplifies the nonlinear eigensolve process and it enforces the correct radiation condition, leading to far more trustworthy solutions to our bent waveguide problem.
    \item More information on the implementation of a PML for solving PDEs numerically is found in B\'{e}renger's publication \cite{berenger1994perfectly} for finite difference methods and in~\cite{collino1998perfectly, Matuszyk_Demkowicz_13, PardoDemkowiczTorres-VerdinMichler08} for finite element methods. 
\end{itemize}



\begin{remark}
    \revC{In \cite{collino1998perfectly},} Collino and Monk carried out \revC{an analysis of the impact on} the Hankel function's asymptotic behavior~\eqref{eq:hankel_asymptotics}. 
    \revC{However, their analysis used} the $H^{(1)}_\beta$ Hankel functions rather than $H^{(2)}_\beta$, as well as a PML path with positive imaginary components due to their choice of sign (convention) in the time-harmonic ansatz (which is opposite to the one used in the present work, that coincides with \revC{Marcuse's ansatz} in \cite{marcuse1982light}).
\end{remark}

\begin{remark}\label{remark:OrderOfTransformations}
    It is important to emphasize that $z=\zpml$ replaces the physical coordinate $r=r_0 + b$, and not the transformed coordinate $x_2=b_2$ defined in \eqref{eq:sol3laycirc_rescaled}. 
    Only after transforming $r \mapsto z(r)$ in the PML region is the transformation made to $x_2$: 
    \[
        x_2 \mapsfrom (r_0 + a) \ln\left( \frac{ z(r) }{ r_0 + a } \right) \, .
    \]
    Thus, the power series at the end of the right cladding region (with argument $x_2$) must be evaluated at
    \[
        b_2 \mapsfrom b_{2,\text{PML}} = (r_0 + a) \ln\left( \frac{ \zpml }{ r_0 + a } \right) \, .
    \]
    The complex-valued natural logarithm is evaluated in its principal branch so that its imaginary part is in the interval $[-\pi, \pi)$. \revC{The outer boundary condition operator is in this case defined by
    $$
    \text{BC}_{\text{outer}} = C_2 \Bfirst^{\rclad}_{\mu_2}(b_{2,\text{PML}}) + D_2 \Bsecond^{\rclad}_{\mu_2}(b_{2,\text{PML}}).
    $$
    Since $b_{2,\text{PML}}$ is not a real number whenever $\pmlparameter\ne 0$, Algorithms \ref{algo:bessel_frobenius} and \ref{algo:bessel_frobenius_dmu} must allow input $x$ to be complex-valued.}
\end{remark}

\subsection{Numerical experiments using the PML BC}
Using the same conditions and waveguide properties as were utilized in Section~\ref{sec:3layer_circ_eig}, the eigensolutions are computed using the PML BC \eqref{eq:PML_bc}, rather than the impedance BC, as the bend radius is varied. 
We set the end of the boundary layer to be the point $\zpml$ defined by \eqref{eq:zpml_definition},
using a PML strength of $\pmlparameter = 800$. For completeness, we mention that the PML path is assumed to begin at the midpoint between the core-cladding interface and the right boundary; that choice, however, does not impact the eigenvalue solver, and is not relevant to this section's graphs because the mode profile plots are traced along the real axis.

\begin{figure}[htb]
    \begin{subfigure}{0.33\textwidth}
        \includegraphics[width=\linewidth]{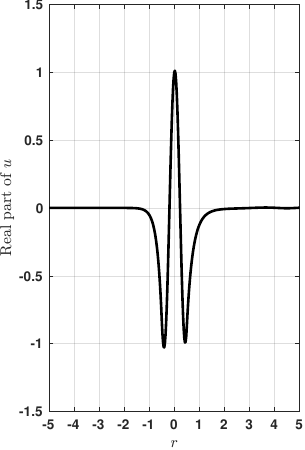}
        \caption{$r_0 = 10400$}
        \label{plot:2ndEvenMode-PMLBC-r0=10400}
    \end{subfigure}%
    \begin{subfigure}{0.33\textwidth}
        \includegraphics[width=\linewidth]{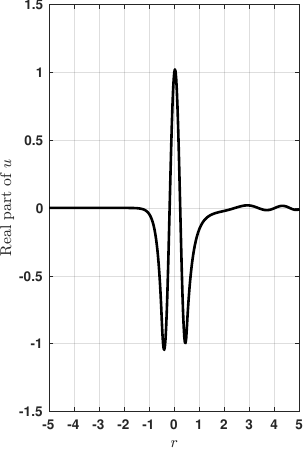}
        \caption{$r_0 = 7800$}
        \label{plot:2ndEvenMode-PMLBC-r0=7800}
    \end{subfigure}%
    \begin{subfigure}{0.33\textwidth}
        \includegraphics[width=\linewidth]{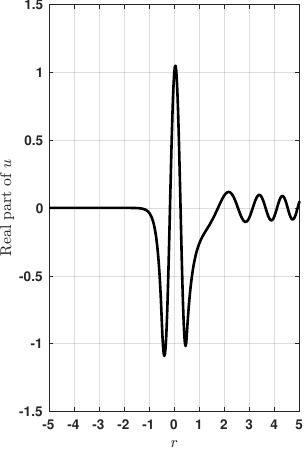}
        \caption{$r_0 = 5200$}
        \label{plot:2ndEvenMode-PMLBC-r0=5200}
    \end{subfigure}%
    \caption{Predicted real component of the second even mode's profile in the three-layer coiled waveguide using a PML BC for different bend radii. The corresponding eigenvalue information is listed in Table~\ref{tab:eigval3_vs_r0-pml}. Compare these mode profiles with those shown in Fig.~\ref{fig:mode3_vs_r0-imp}, which are computed using an impedance BC, to observe the differences in the features of the radiating oscillations toward the right boundary of the waveguide.}
    \label{fig:mode3_vs_r0-pml}
\end{figure}

\begin{table}[htb]
    \centering
    \caption{Computed eigenvalues, using a PML BC, of the second even mode as the coiling radius is tightened. Figure~\ref{fig:mode3_vs_r0-pml} depicts their corresponding mode profiles. Compare these model-predicted mode eigenvalues with those delineated in Table~\ref{tab:eigval3_vs_r0-imp}, which were computed using an impedance BC: the digits, or entire values, that differ are shaded.}
    \label{tab:eigval3_vs_r0-pml}
    {\small
    \begin{tabular}{|p{1.0cm}|p{0.38cm}|p{2.95cm}|p{2.95cm}|p{2.95cm}|}
        \hline
        $r_0$ & \mbox{} & $\lambda$ & $\beta=\sqrt{\lambda}$ & $\sqrt{\mu}=\beta/r_0$ \rule[-1.5mm]{0mm}{6mm} \\
        \hline
        \multirow{2}{1.0cm}{$10400$} & ${\Re}$ & \scalebox{0.75}{$5.11869686\hlgray{447172} \cdot 10^{12}$} & \scalebox{0.75}{$2.26245372\hlgray{648187} \cdot 10^{6}$} & \scalebox{0.75}{$2.17543627\hlgray{546334}\cdot 10^{2}$} \\
        \mbox{} & ${\Im}$ & \scalebox{0.75}{\hlgray{$-3.59916299495178 \cdot 10^{3}$}} & \scalebox{0.75}{\hlgray{$-7.95411405065176 \cdot 10^{-4}$}} & \scalebox{0.75}{$\hlgray{-7.64818658716516} \cdot 10^{-8}$} \\
        \hline
        \multirow{2}{1.0cm}{$7800$} & ${\Re}$ & \scalebox{0.75}{$2.879271\hlgray{68048203} \cdot 10^{12}$} & \scalebox{0.75}{$1.696841678\hlgray{08374} \cdot 10^{6}$} & \scalebox{0.75}{$2.1754380\hlgray{488253076}\cdot 10^{2}$} \\
        \mbox{} & ${\Im}$ & \scalebox{0.75}{\hlgray{$-1.00373251044341 \cdot 10^{5}$}} & \scalebox{0.75}{\hlgray{$-0.0295764927101785$}} & \scalebox{0.75}{\hlgray{$-3.79185803976647 \cdot 10^{-6}$}} \\
        \hline
        \multirow{2}{1.0cm}{$5200$} & ${\Re}$ & \scalebox{0.75}{$1.2796\hlgray{8375031025} \cdot 10^{12}$} & \scalebox{0.75}{$1.1312\hlgray{3107732720} \cdot 10^{6}$} & \scalebox{0.75}{$2.175\hlgray{44437947539}\cdot 10^{2}$} \\
        \mbox{} & ${\Im}$ & \scalebox{0.75}{$-\hlgray{1.76816227029980} \cdot 10^{6}$} & \scalebox{0.75}{$-0.\hlgray{781521258449466}$} & \scalebox{0.75}{$-1.\hlgray{5029254970182} \cdot 10^{-4}$} \\
        \hline
    \end{tabular}
    }
\end{table}

Fig.~\ref{fig:mode3_vs_r0-pml} displays the predicted second even mode profiles under three distinct bend radii; the corresponding eigenvalue results are tabulated in Table~\ref{tab:eigval3_vs_r0-pml}.
Note the differences in the \revC{period and amplitude of} the outgoing oscillations observed in the predicted mode profile (particularly big in the $r_0 = 5200$ case), when using the PML BC (Fig.~\ref{fig:mode3_vs_r0-pml}) in comparison with \revC{the corresponding profiles obtained with the} impedance BC (Fig.~\ref{fig:mode3_vs_r0-imp}). \revC{Despite these considerable differences,} the computed eigenvalues are still quite similar to one another (all of the imaginary components are \revC{of} the same order of magnitude, cf.~Table~\ref{tab:eigval3_vs_r0-imp} vs.\ Table~\ref{tab:eigval3_vs_r0-pml}).

We continue the comparison of the results obtained with both BC settings by observing the propagation constant, $\beta$, of all the propagating modes at different bend radii.
Moreover, we add a smaller value for $r_0$, which does not lead to a successful convergence with the impedance BC but does deliver a satisfactory result under the PML approach. Table~\ref{tab:eigval3_comparison_vs_r0} contains all the resulting values of $\beta$, comparing \revB{the} 
impedance BC and the PML approach, for the two even modes and the odd mode \revC{at} four bend radii. 
\revC{Particularly notable in these results} is that the real part of $\beta$ varies \revC{almost} linearly with $r_0$, but the imaginary part grows (in absolute value) \revC{nearly} exponentially as $r_0$ decreases. Apart from that observation, an important takeaway is that at the smallest $r_0$ the value of $-\Im(\beta)$ of the second even mode induces a $99.99\%$ loss in about one radian. The odd mode, in turn, requires a few dozen coils to fully vanish. The first mode, however, even at the tightest bend, exhibits a loss factor that needs to act for nearly a million coils to make it vanish.

\begin{table}[htb]
    \centering
    \caption{Comparison of the model-predicted mode propagation constants in the three-layer circularly bent waveguide for \revC{four} distinct bend radii, using either the impedance BC or the PML BC, applied at $r = r_0 + b$. The digits, or entire values, that differ between the two conditions are shaded in the impedance BC column only. The real components tend to be more aligned than the imaginary components, and, at least for this specific problem, are of the same order of magnitude. 
    Moreover, the imaginary components are better matched between these two BCs for weaker bends (i.e.\ larger bend radii). 
    Because the PML BC is a better representation of the proper radiation condition, its results are trusted to be more accurate than those found using the impedance BC. 
    }
    \label{tab:eigval3_comparison_vs_r0}
    \begin{tabular}{|p{1.5cm}|p{1.1cm}|p{0.6cm}|p{3.9cm}|p{3.3cm}|}
        \hline
        Mode & $r_0$ & \mbox{} & $\beta$ (using impedance BC) & $\beta$ (using PML BC) \\
        \hline
        \multirow{2}{1.5cm}{$1^{\text{st}}$ Even} & \multirow{2}{1.1cm}{$10400$} & ${\Re}$ & 
        \scalebox{0.75}{$2.26362060047958 \cdot 10^{6}$} & 
        \scalebox{0.75}{$2.26362060047958 \cdot 10^{6}$} \\
        \mbox{} & \mbox{} & ${\Im}$ & 
        \scalebox{0.75}{\hlgray{$-1.02194399501288 \cdot 10^{-26}$}} & 
        \scalebox{0.75}{{$-2.63161591219032 \cdot 10^{-30}$}} \\
        \hline
        \multirow{2}{1.5cm}{$1^{\text{st}}$ Even} & \multirow{2}{1.1cm}{$7800$} & ${\Re}$ & 
        \scalebox{0.75}{$1.69771848771636 \cdot 10^{6}$} & 
        \scalebox{0.75}{$1.69771848771636 \cdot 10^{6}$} \\
        \mbox{} & \mbox{} & ${\Im}$ & 
        \scalebox{0.75}{\hlgray{$-3.55201875479254 \cdot 10^{-24}$}} & 
        \scalebox{0.75}{{$-7.37577942903455 \cdot 10^{-25}$}} \\
        \hline
        \multirow{2}{1.5cm}{$1^{\text{st}}$ Even} & \multirow{2}{1.1cm}{$5200$} & ${\Re}$ & 
        \scalebox{0.75}{$1.1318\hlgray{4377894959} \cdot 10^{6}$} & 
        \scalebox{0.75}{$1.1318{1802321074} \cdot 10^{6}$} \\
        \mbox{} & \mbox{} & ${\Im}$ & 
        \scalebox{0.75}{\hlgray{$-1.00005695603398$}} & 
        \scalebox{0.75}{{$-2.04607567975992 \cdot 10^{-15}$}} \\
        \hline
        \multirow{2}{1.5cm}{$1^{\text{st}}$ Even} & \multirow{2}{1.1cm}{$2600$} & ${\Re}$ & 
        \scalebox{0.75}{N.A.} & 
        \scalebox{0.75}{$5.65923463817321 \cdot 10^{5}$} \\
        \mbox{} & \mbox{} & ${\Im}$ & 
        \scalebox{0.75}{N.A.} & 
        \scalebox{0.75}{$-3.21177027104337 \cdot 10^{-6}$} \\
        \hline
        \multirow{2}{1.5cm}{$1^{\text{st}}$ Odd} & \multirow{2}{1.1cm}{$10400$} & ${\Re}$ & 
        \scalebox{0.75}{$2.26315767840190 \cdot 10^{6}$} & 
        \scalebox{0.75}{$2.26315767840190 \cdot 10^{6}$} \\
        \mbox{} & \mbox{} & ${\Im}$ & 
        \scalebox{0.75}{$-\hlgray{4.92656786477296} \cdot 10^{-20}$} & 
        \scalebox{0.75}{$-{5.66184601060354} \cdot 10^{-20}$} \\
        \hline
        \multirow{2}{1.5cm}{$1^{\text{st}}$ Odd} & \multirow{2}{1.1cm}{$7800$} & ${\Re}$ & 
        \scalebox{0.75}{$1.69736779822896 \cdot 10^{6}$} & 
        \scalebox{0.75}{$1.69736779822896 \cdot 10^{6}$} \\
        \mbox{} & \mbox{} & ${\Im}$ & 
        \scalebox{0.75}{\hlgray{$-8.02404417111743 \cdot 10^{-16}$}} & 
        \scalebox{0.75}{{$-4.97996447610167 \cdot 10^{-14}$}} \\
        \hline
        \multirow{2}{1.5cm}{$1^{\text{st}}$ Odd} & \multirow{2}{1.1cm}{$5200$} & ${\Re}$ & 
        \scalebox{0.75}{$1.13157775618\hlgray{699} \cdot 10^{6}$} & 
        \scalebox{0.75}{$1.13157775618{741} \cdot 10^{6}$} \\
        \mbox{} & \mbox{} & ${\Im}$ & 
        \scalebox{0.75}{\hlgray{$-1.51030428594905 \cdot 10^{-7}$}} & 
        \scalebox{0.75}{{$-3.72804455077520 \cdot 10^{-8}$}} \\
        \hline
        \multirow{2}{1.5cm}{$1^{\text{st}}$ Odd} & \multirow{2}{1.1cm}{$2600$} & ${\Re}$ & 
        \scalebox{0.75}{N.A.} & 
        \scalebox{0.75}{$5.65787956064918 \cdot 10^{5}$} \\
        \mbox{} & \mbox{} & ${\Im}$ & 
        \scalebox{0.75}{N.A.} & 
        \scalebox{0.75}{$-0.0159239556531208$} \\
        \hline
        \multirow{2}{1.5cm}{$2^{\text{nd}}$ Even} & \multirow{2}{1.1cm}{$10400$} & ${\Re}$ & 
        \scalebox{0.75}{$2.26245372\hlgray{707224} \cdot 10^{6}$} & 
        \scalebox{0.75}{$2.26245372{648187} \cdot 10^{6}$} \\
        \mbox{} & \mbox{} & ${\Im}$ & 
        \scalebox{0.75}{\hlgray{$-2.68135826135536 \cdot 10^{-5}$}} & 
        \scalebox{0.75}{{$-7.95411405065176 \cdot 10^{-4}$}} \\
        \hline
        \multirow{2}{1.5cm}{$2^{\text{nd}}$ Even} & \multirow{2}{1.1cm}{$7800$} & ${\Re}$ & 
        \scalebox{0.75}{$1.6968416\hlgray{4784584} \cdot 10^{6}$} & 
        \scalebox{0.75}{$1.6968416{7808374} \cdot 10^{6}$} \\
        \mbox{} & \mbox{} & ${\Im}$ & 
        \scalebox{0.75}{\hlgray{$-1.69915206327735 \cdot 10^{-3}$}} & 
        \scalebox{0.75}{{$-0.0295764927101785$}} \\
        \hline
        \multirow{2}{1.5cm}{$2^{\text{nd}}$ Even} & \multirow{2}{1.1cm}{$5200$} & ${\Re}$ & 
        \scalebox{0.75}{$1.1312\hlgray{0463654890} \cdot 10^{6}$} & 
        \scalebox{0.75}{$1.1312{3107732720} \cdot 10^{6}$} \\
        \mbox{} & \mbox{} & ${\Im}$ & 
        \scalebox{0.75}{$-0.\hlgray{994847657526836}$} & 
        \scalebox{0.75}{$-0.{781521258449466}$} \\
        \hline
        \multirow{2}{1.5cm}{$2^{\text{nd}}$ Even} & \multirow{2}{1.1cm}{$2600$} & ${\Re}$ & 
        \scalebox{0.75}{N.A.} & 
        \scalebox{0.75}{$5.65620469836942 \cdot 10^{5}$} \\
        \mbox{} & \mbox{} & ${\Im}$ & 
        \scalebox{0.75}{N.A.} & 
        \scalebox{0.75}{$-8.96795892357474$} \\
        \hline
    \end{tabular}
\end{table}

\begin{figure}[htb]
    \centering
    \begin{subfigure}{0.323\textwidth}
        \includegraphics[width=\linewidth]{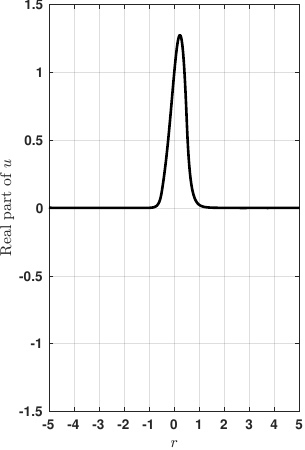}
        \caption{First even mode}
        \label{plot:1stEvenMode_r=2600_PML}
    \end{subfigure}
    \begin{subfigure}{0.323\textwidth}
        \includegraphics[width=\linewidth]{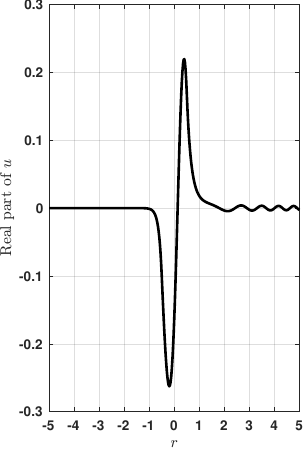}
        \caption{First odd mode}
        \label{plot:OddMode_r=2600_PML}
    \end{subfigure}
    \begin{subfigure}{0.323\textwidth}
        \includegraphics[width=\linewidth]{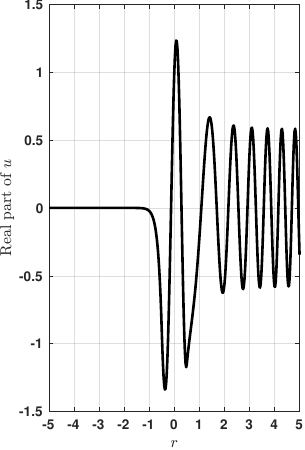}
        \caption{Second even mode}
        \label{plot:2ndEvenMode_r=2600_PML}
    \end{subfigure}
    \caption{Predicted mode profiles of the three-layer bent waveguide using the PML BC and a bend radius of $r_0 = 2600$. All modes exhibit noticeable distortion effects from the circular coiling.}
    \label{fig:modes_r2600}
\end{figure}

Unlike with larger values of $r_0$, at the tightest bend corresponding to $r_0 = 2600$, it is possible to clearly visualize the coiling-induced distortion of the mode profile ``leaning'' toward the outside of the bend in all three modes (see Fig.~\ref{fig:modes_r2600}). 
Looking at the location of the peak of \revC{the} even mode's graph (Fig.~\ref{plot:1stEvenMode_r=2600_PML}), one can ascertain that the mode is leaning to the right, as expected. 
Furthermore, the odd mode's profile (Fig.~\ref{plot:OddMode_r=2600_PML}) exhibits visible fluctuations toward the outer boundary---an indication that the mode is leaking energy and may no longer be normalizable because the bending transformed it from a perfectly guided mode into a \textit{leaky mode}. 
Finally, the second even mode at $r_0=2600$ (Fig.~\ref{plot:2ndEvenMode_r=2600_PML}) continues the trend of increasingly stronger outgoing oscillations already seen with $r_0=10400$, $7800$, and $5200$ (in Fig.~\ref{fig:mode3_vs_r0-pml}a, b and c, respectively), now with a noticeably higher amplitude and frequency.

\subsection{Main differences between the results with the impedance BC and the PML}
A central takeaway of the results presented to this point is that the impedance BC \revC{produces results that differ significantly from those obtained with the PML.} We have already mentioned the fact that the impedance BC may be inappropriate whenever the direction of the wave incidence to the boundary is far from normal. 
\revC{The following observations from the numerical results, which are consistent across both implementations, provide additional evidence that the PML offers a more accurate and reliable treatment of the radiation condition.}
\begin{enumerate}
    \item Note the major visual difference for the second even mode profile between \revC{plot~c of Fig.~\ref{fig:mode3_vs_r0-imp} and plot~c of Fig.~\ref{fig:mode3_vs_r0-pml}}. 
    \revC{The oscillation amplitude in the right cladding increases abruptly in the result obtained with the impedance BC (see plots a and b of Fig.~\ref{fig:mode3_vs_r0-pml}). Conversely,} the corresponding PML result displays a coherent transition between the profile at $r_0=7800$ (Fig.~\ref{fig:mode3_vs_r0-pml}b) and the profile at $r_0=2600$ (Fig.~\ref{fig:modes_r2600}c).
    \item \revC{Fig.~\ref{fig:mode3_vs_r0-imp} shows} that the solution profile, after a few fluctuations, nearly vanishes at the right boundary. This \revC{results from} a magnitude disparity between the terms of the impedance BC (cf. \eqref{eq:impBC}, where the coefficient that multiplies $u$ is much larger than the one acting on $d u/d r$). In contrast, the mode profiles obtained with the PML BC do not present that feature (consider Fig.~\ref{fig:mode3_vs_r0-pml} and Fig.~\ref{fig:modes_r2600}).
    \item In Table~\ref{tab:eigval3_comparison_vs_r0}, we can observe how $\Im(\beta)$ varies as $r_0$ decreases, and that the range of orders of magnitude through which it \revC{varies} depends on the mode.
    \revC{Note} that, for the first even mode, $\Im(\beta)$ has values in the $10^{-30}$ to $10^{-20}$ range for the two largest bend radii but, in the impedance BC case, at $r_0=5200$ we get a value completely out of trend ($\sim 10^0$), whereas the PML result at $r_0=5200$ \revC{shows} a good alignment with the former values, and this result is accompanied by the one at $r_0=2600$, \revC{confirming the same pattern}.
    \item The fact that \revC{the PML formulation converges more readily than the impedance-BC formulation} to the sought unknowns, even for a smaller bend radius, supports the \revC{claim that} the PML's complex-path tracing \revC{is} a more appropriate or natural approximation to the outgoing radiation phenomenon.
\end{enumerate}

\subsection{Robustness of the predicted solution with respect to the PML parameters}
We additionally study the effect of changing the parameters of the PML BC (i.e., the location of $\zpml$ in both directions of the complex plane) on the computed propagation constant $\beta$. 
To this end, we fix the bend radius at $r_0 = 5200$ and, again focusing on the second even mode, assess how the mode's propagation constant changes as a function of the PML strength $\pmlparameter$ and the size of the layer (by modifying the real part of $\zpml$). 
The results of this study are listed in Table~\ref{tab:eigval3_vs_pml_parameter} for a fixed layer length and seven values of $\pmlparameter$, and in Table~\ref{tab:eigval3_vs_zpml_location} for a fixed PML strength parameter and four different layer lengths. 
These results show that the predicted eigenvalue behaves robustly with regard to the PML parameters, since its value has \revC{only} small variations over a wide range of the PML parameters. While it is much less sensitive to the size of the PML, it does require a sufficiently large PML strength parameter, $\pmlparameter$, to become steady. However, if $\pmlparameter$ grows too much, the computations are somewhat slower. On the other hand, even though several of the PML lengths return the same 15 digits of $\beta$, for the sake of simplicity and symmetry, using the original $\zpml$ defined in \eqref{eq:zpml_definition} is recommended.
Consequently, we consider the case with $\pmlparameter=800$ and $\zpml= r_0 + b - i \, \pmlparameter / \wavenum_1$ (see fifth row of Table~\ref{tab:eigval3_vs_pml_parameter} and first row of Table~\ref{tab:eigval3_vs_zpml_location}, respectively) as the reference solution against which the remaining results are compared, particularly because $\pmlparameter=800$ is the smallest value at which the first fifteen digits in both $\Re(\beta)$ and $\Im(\beta)$ \revC{become} fixed.

\begin{table}[htb]
    \centering
    \caption{Predicted eigenvalues of the second even mode for a fixed bend radius of $r_0 = 5200$, using the PML BC applied at $\zpml = r_0 + b - i\, \pmlparameter / \wavenum_1$, as a function of the parameter $\pmlparameter$. 
    We highlight in \hlgray{gray} all the digits that differ from the reference solution: $\pmlparameter = 800$ case. The computational speed noticeably decreased as $\pmlparameter$ increased, especially for the $\pmlparameter = 1600$ and $\pmlparameter = 3200$ cases.}
    \label{tab:eigval3_vs_pml_parameter}
    \begin{tabular}{|p{2.9cm}|p{0.5cm}|p{3.7cm}|}
        \hline
        $r_0 = 5200$ & \mbox{} & $\beta$ \revC{(using PML BC)} \\
        \hline
        \multirow{2}{2.5cm}{$\pmlparameter = 50$} & ${\Re}$ & \scalebox{0.75}{$0.1131231\hlgray{11157010} \cdot 10^{7}$} \\
        \mbox{} & ${\Im}$ & \scalebox{0.75}{$-0.7\hlgray{65959119625596}$} \\
        \hline
        \multirow{2}{2.5cm}{$\pmlparameter = 100$} & ${\Re}$ & \scalebox{0.75}{$0.113123107\hlgray{805648} \cdot 10^{7}$} \\
        \mbox{} & ${\Im}$ & \scalebox{0.75}{$-0.781\hlgray{048100766258}$} \\
        \hline
        \multirow{2}{2.5cm}{$\pmlparameter = 200$} & ${\Re}$ & \scalebox{0.75}{$0.1131231077327\hlgray{33} \cdot 10^{7}$} \\
        \mbox{} & ${\Im}$ & \scalebox{0.75}{$-0.78152\hlgray{0834783452}$} \\
        \hline
        \multirow{2}{2.5cm}{$\pmlparameter = 400$} & ${\Re}$ & \scalebox{0.75}{$0.113123107732720 \cdot 10^{7}$} \\
        \mbox{} & ${\Im}$ & \scalebox{0.75}{$-0.781521258449\hlgray{540}$} \\
        \hline
        \multirow{2}{2.9cm}{$\pmlparameter = 800$} & ${\Re}$ & \scalebox{0.75}{$0.113123107732720 \cdot 10^{7}$} \\
        \mbox{} & ${\Im}$ & \scalebox{0.75}{$-0.781521258449466$} \\
        \hline
        \multirow{2}{2.9cm}{$\pmlparameter = 1600$} & ${\Re}$ & \scalebox{0.75}{$0.113123107732720 \cdot 10^{7}$} \\
        \mbox{} & ${\Im}$ & \scalebox{0.75}{$-0.781521258449466$} \\
        \hline
        \multirow{2}{2.9cm}{$\pmlparameter = 3200$} & ${\Re}$ & \scalebox{0.75}{$0.113123107732720 \cdot 10^{7}$} \\
        \mbox{} & ${\Im}$ & \scalebox{0.75}{$-0.781521258449466$} \\
        \hline
    \end{tabular}
\end{table}

\begin{table}[htb]
    \centering
    \caption{Computed eigenvalues of the second even mode for a fixed bend radius, $r_0 = 5200$, and PML strength, $\pmlparameter = 800$, using a PML BC with the boundary layers ending at different locations of $\zpml$. All of the digits that differ from the reference solution (first row) are highlighted in \hlgray{gray}.}
    \label{tab:eigval3_vs_zpml_location}
    \begin{tabular}{|p{6.5cm}|p{0.5cm}|p{3.75cm}|}
        \hline
        $r_0 = 5200$, $\pmlparameter = 800$ & \mbox{} & $\beta$ \revC{(using PML BC)} \\
        \hline
        \multirow{2}{6.5cm}{$\zpml = r_0 + b - i \, \pmlparameter / \wavenum_1$} & 
            ${\Re}$ & 
            \scalebox{0.75}{$0.113123107732720 \cdot 10^{7}$} \\
        \mbox{} & ${\Im}$ & \scalebox{0.75}{$-0.781521258449466$} \\
        \hline
        \multirow{2}{6.5cm}{$\zpml = r_0 + a + 0.75 (b - a) -i  \, \pmlparameter / \wavenum_1$} & 
            ${\Re}$ & 
            \scalebox{0.75}{$0.113123107732720 \cdot 10^{7}$} \\
        \mbox{} & ${\Im}$ & \scalebox{0.75}{$-0.781521258449466$} \\
        \hline
        \multirow{2}{6.5cm}{$\zpml = r_0 + a + 0.50 (b - a) - i \, \pmlparameter / \wavenum_1$} & 
            ${\Re}$ & 
            \scalebox{0.75}{$0.113123107732720 \cdot 10^{7}$} \\
        \mbox{} & ${\Im}$ & \scalebox{0.75}{$-0.781521258449466$} \\
        \hline
        \multirow{2}{6.5cm}{$\zpml = r_0 + a + 0.25 (b - a) - i \, \pmlparameter / \wavenum_1$} & 
            ${\Re}$ & 
            \scalebox{0.75}{$0.113123107732720 \cdot 10^{7}$} \\
        \mbox{} & ${\Im}$ & \scalebox{0.75}{$-0.7815212584494\hlgray{55}$} \\
        \hline
    \end{tabular}
\end{table}

\section{Comparison with loss formulas for the bent slab waveguide}
\label{sec:BendLoss}

Finally, we compare our results with values obtained from mode bend loss formulas that are available in the literature; these algebraic formulas provide the loss factor per radian, corresponding to the $-\Im(\beta)$ values in our notation. 
The comparison includes results from three sources relevant to our methodology \cite{marcatili1969bends, marcuse1982light, takuma1981bent}, which have already been briefly described in \revC{Section}~\ref{sec:introduction}. All three contributions account only for geometric effects and provide closed formulas that are easy to implement.

\revC{Marcatili's formulas} \cite{marcatili1969bends} are ultimately applicable only to the fundamental transverse mode of the waveguide, but with two different polarization directions. We use the formula~(19) thereof, which best coincides with our problem setup. 
As the second reference method, we make use of equation (9.6-27) of Marcuse's book~\cite{marcuse1982light}. That formula can be applied to any mode as long as we know the real propagation constants \revC{of the modes of the straight slab waveguide}. 
The third reference formula corresponds to equation (43) of Takuma et al.\ \cite{takuma1981bent}. This is applicable to any mode, $n = 0, 1, 2{\ldots}$, but the authors acknowledge the inaccuracy of their approach \revC{in capturing} the propagation constants (eigenvalues) \revC{of} higher-order modes ($n \geq 1$).

\begin{figure}[htb]
    \centering
    \begin{subfigure}{0.32\columnwidth}
        \includegraphics[width=\textwidth]{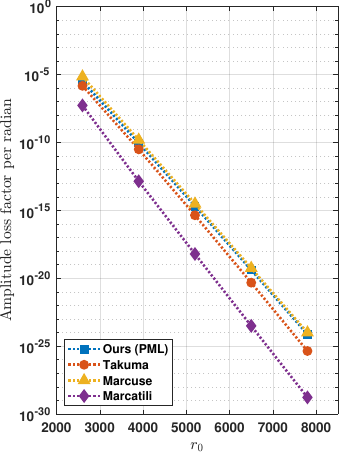}
        \caption{First even mode}
        \label{fig:even1-loss-comparison}
    \end{subfigure} \hfill %
    \begin{subfigure}{0.32\columnwidth}
        \includegraphics[width=\textwidth]{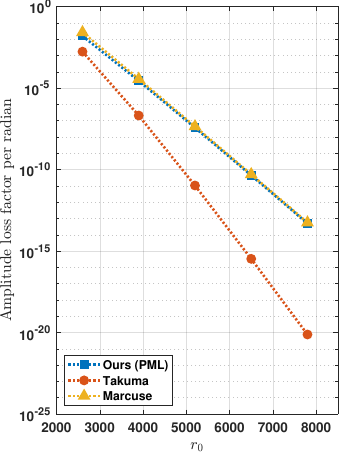}
        \caption{First odd mode}
        \label{fig:odd1-loss-comparison}
    \end{subfigure} \hfill %
    \begin{subfigure}{0.32\columnwidth}
        \includegraphics[width=\textwidth]{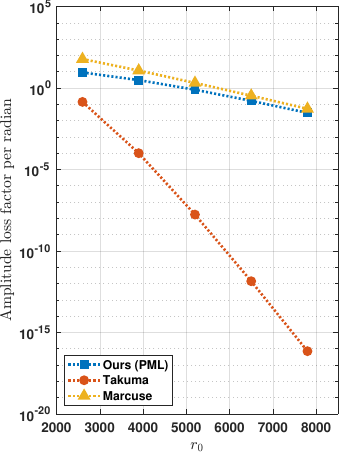}
        \caption{Second even mode}
        \label{fig:even2-loss-comparison}
    \end{subfigure}
    \caption{Comparison of the circularly curved three-layer waveguide's amplitude loss factors, $-{\Im}(\beta)$, as a function of the bend radius ($r_0$) for the (a)~first even mode, (b)~first odd mode, and (c)~second even mode. Results obtained with our numerical approach using the PML BC are compared with loss values obtained with formulas from the literature by Marcatili \cite{marcatili1969bends}, Marcuse \cite{marcuse1982light}, and Takuma et al.~\cite{takuma1981bent}.}
    \label{fig:loss-comparison}
\end{figure}

We compute the mode bend loss estimates from these three formulas and compare them with the values computed with our methodology for the first three modes of the waveguide using five different bend radii between 2600 and 7800.
Fig.~\ref{fig:loss-comparison} depicts the results of this comparison; note that we show results from Marcatili's formula only for the fundamental mode (i.e., the first even mode) where this formula is applicable (Fig.~\ref{fig:even1-loss-comparison}). 
We observe that in all cases, Marcuse's formula remains close to our results throughout the entire range of bend radii, \revC{matching our results} to at least the same order of magnitude. Marcatili's formula \revC{deviates the most from our results} for the first even mode. Takuma's formula is within one order of magnitude for this fundamental mode but strongly diverges from our results for the other two modes.

As expected, the losses decrease as the bend radius increases, and the higher-order modes (i.e., the first odd mode and the second even mode) exhibit higher rates of loss than the fundamental mode. 
The effect of higher-order modes having higher bend losses than their lower-order mode counterparts is called \textit{differential mode loss}~\cite[{\S}5C]{richardson2010high}, and is often used to filter out unwanted higher-order modes from the waveguide output.

\revA{Despite the good agreement between Marcuse's formula and our results (especially for larger bend radii), our methodology can retain its accuracy and effectiveness as $r_0$ becomes smaller, beyond the range of applicability of Marcuse's approximations. 
Moreover, as discussed in Section~\ref{sec:introduction}, the estimation of bend losses is just one practical output of our methodology; it can additionally deliver the eigenfunction profiles (as shown in Sections \ref{sec:3layer_circ_eig} and \ref{sec:PML-BC}), which none of the other references offer.}



\section{Conclusions}

\revC{
We have presented a study of circularly coiled waveguides with parameters representative of standard step-index optical fibers. 
This leads to an eigenproblem with complex-valued eigenvalues that determine the orders of the corresponding eigenfunctions, which are expressed in terms of Bessel functions. 
The associated Bessel functions also have complex-valued arguments of large magnitude.
Such functions are difficult to evaluate reliably and accurately using standard numerical approaches. 
In Section~\ref{sec:EvalBesselFunctions}, we demonstrated that, by applying the conformal mapping~\eqref{eq:MappingTox} to the independent variable and using the classical Frobenius method, one obtains recursion relations that enable the efficient evaluation of these Bessel functions for large arguments and complex-valued orders. 
Combined with high-precision floating-point arithmetic, this approach provides the numerical accuracy required for the parameter regimes considered here. 
Appendix~\ref{sec:convergence} further illustrates the convergence behavior of the proposed algorithms.
Furthermore, through Algorithm~\ref{algo:eigensolver-3layer}, we developed a methodology for solving the bent-waveguide eigenproblem by using the corresponding straight-waveguide solutions to construct effective initial guesses for the nonlinear eigensolver.
}

\revC{
In Section~\ref{sec:PML-BC}, we replaced the impedance BC imposed on the outer side of the bend with a PML that provides a more reliable representation of the radiation condition for the eigenvalue problem. 
Our numerical experiments show that the PML formulation can produce results differing by several orders of magnitude from those obtained with the impedance BC, particularly in the imaginary part of the propagation constant, which determines the confinement loss of the mode, and for tighter bends (i.e.\ smaller values of the bend radius $r_0$). 
These results demonstrate that the choice of radiation condition can have a substantial impact on the predicted bend losses.
}

\revC{
Finally, Section~\ref{sec:BendLoss} demonstrates a practical application of the proposed methodology by using the PML-based eigensolutions to estimate bend losses in a slab waveguide and comparing them with approximate formulas available in the literature.
The resulting eigenvalues and eigenfunctions provide high-accuracy reference solutions that can be used to verify and benchmark numerical models of bent waveguides. 
These solutions have already been used in our three-dimensional Maxwell finite element model of the bent waveguide~\cite{mora2026bent} and can support further verification studies in future work.
}

\subsection*{Source code and reproducibility}
\revC{Readers are encouraged to visit our open-source repository at \url{https://github.com/cgt3/WaveguideSolver} to access and use the Julia implementation of our Bessel function evaluation algorithms, obtain the modes and eigenvalues of the bent waveguides studied here with the provided scripts, and reproduce the main results of this paper.}

\section*{Disclaimers}
This article has been approved for public release; distribution unlimited. 
Public Affairs release approval {\#}AFRL-2025-5370. 
The views expressed in this article are those of the authors and do not necessarily reflect the official policy or position of the Department of the Air Force, of the Department of Defense, nor of the U.S. government.
%
\bibliographystyle{siamplain}
\bibliography{bib/ref_paper}
%
\appendix
\section{Numerical evidence for the spectral analysis of a circular waveguide with impedance BC}
\label{sec:NumericalGlazmanCriterion}
In~\cite{Demkowicz_Gopalakrishnan_Heuer_24}, the authors studied the well-posedness and stability of a straight acoustical waveguide with impedance BC.
Using separation of variables, the solution is expanded into a set of modes (eigenfunctions). 
The non-triviality of the problem lies in the fact that---contrary to the homogeneous waveguide with sound-hard BC~\cite{melenk2025waveguide1}---the operator that defines the modes here is \emph{not} self-adjoint; hence the modes are orthogonal neither in the $L^2$ nor in the $H^1$ inner product. 
Thus, the completeness of the modes in the $L^2$ and $H^1$ spaces no longer follows from the Sturm--Liouville theory (cf.\ Spectral Theorem for Self-Adjoint Operators). 
Rather, non-self-adjoint operator theory must be employed, as discussed in reference~\cite{Gohberg_Krein_65}. 
The proof of well-posedness relied upon the Glazman Theorem
 (see p.~328 of \cite{Gohberg_Krein_65}), which requires the verification of the Glazman condition expressed in terms of the eigenvalues, $\lambda_n$: 
 \be
    \sum_{\substack{i, j = 1 \\ i \ne j}}^{\infty} 
    \frac{ \Im (\lambda_i)\, \Im (\lambda_j) }{ \vert \lambda_i - \lambda_j \vert^2 } < \infty \, .
    \label{eq:Glazman_condition}
 \ee
 Note that this condition is trivially satisfied for the waveguide problem with hard or soft BCs for which the operator is self-adjoint, which enforces real-valued eigenvalues; hence $\Im(\lambda_j) = 0,\ \forall j$. 
 However, for the straight waveguide with impedance BC, the eigenproblem leads to a transcendental equation involving the tangent function with complex variables,
 which can be manipulated to study the behavior of the equation's roots (i.e., the problem's eigenvalues) and thereby the satisfaction of the Glazman criterion~\eqref{eq:Glazman_condition}. 

With the Bessel eigensolver in hand, we consider the homogeneous circular waveguide ($n(r) = 1$), with a Neumann BC on the inner boundary, \revB{$du/dr(r_0-b)=0$}, and the impedance BC \eqref{eq:impBC} on the outer boundary.
We numerically study the behavior of the first 30 eigenvalues of~\eqref{eq:Bessel_eqn_new}. Since the Bessel eigenvalue equation with impedance BC can be regarded as a non-self-adjoint operator, we computationally check whether the Glazman criterion is also satisfied for the case of a bent waveguide.

The reported experiment was our first ``hands-on sanity check'' on the problem before we engaged in more advanced theories \cite{Demkowicz_Halla_Melenk_26}.

\begin{figure}[htb]
    \centering
    \begin{subfigure}{0.321\textwidth}
        \includegraphics[width = \textwidth]
        {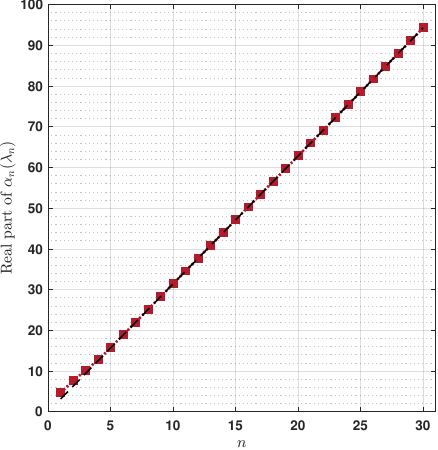}
        \caption{Real component: ${\Re}(\alpha_n)$}
        \label{plot:RealAlphaValues}
    \end{subfigure}
    \begin{subfigure}{0.329\textwidth}
        \includegraphics[width=\textwidth]
        {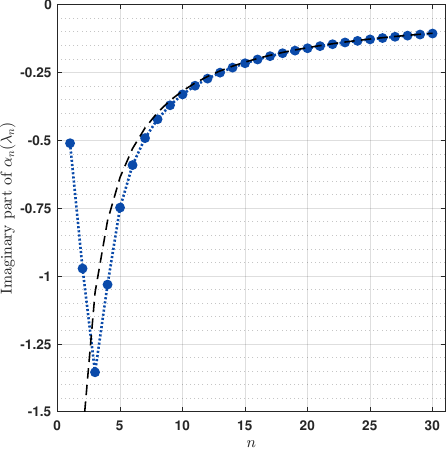}
        \caption{Imaginary component: ${\Im}(\alpha_n)$}
        \label{plot:ImagAlphaValues}
    \end{subfigure}
    \begin{subfigure}{0.324\textwidth}
        \includegraphics[width=\textwidth]
        {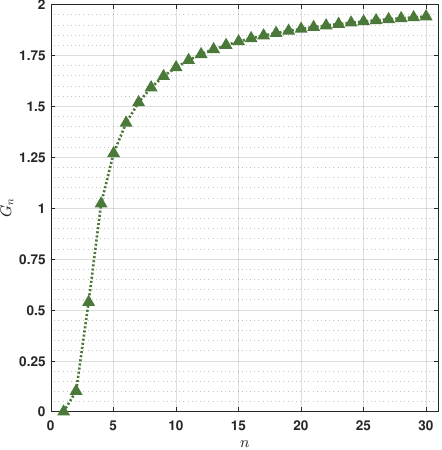}
        \caption{Finite sums: $G_n$}
        \label{plot:GlazmanFiniteSums}
    \end{subfigure}
    \caption{The real (plot~\subref{plot:RealAlphaValues}) and imaginary (plot~\subref{plot:ImagAlphaValues}) components of the $\alpha_{n} = \alpha_{n}(\lambda_{n})$ function~\eqref{eq:AlphaValue} for the first 30 eigenvalues of~\eqref{eq:Bessel_eqn_new} using $n_{0} = 1$, $\wavenum = 10$, $r_0 = 100$, $b = 0.5$, and $d = 1$. According to the theoretical results in \cite{Demkowicz_Gopalakrishnan_Heuer_24}, in the limit of a straight waveguide ($r_0 \to \infty$), $\Re(\alpha_{n}) \to n \pi$ and $\Im(\alpha_n)\to-2bd\kappa/(\pi n)$ (see the black dashed lines in~\subref{plot:RealAlphaValues} and~\subref{plot:ImagAlphaValues}). In plot~\subref{plot:GlazmanFiniteSums}, we present the finite sums $G_n$~\eqref{eq:finite_Glazman} associated with the Glazman condition~\eqref{eq:Glazman_condition}, and observe the convergence of the series. 
    }
    \label{fig:glazman-test}
\end{figure}

Fig.~\ref{fig:glazman-test} depicts the computational results for $n = 1,2,\dots,30$, with parameters $\wavenum = 10$, $r_0 = 100$, $b = 0.5$, and $d = 1$. 
We show the real (plot~\ref{plot:RealAlphaValues}) and imaginary (plot~\ref{plot:ImagAlphaValues}) components of the function 
\be\label{eq:AlphaValue}
    \alpha_n(\lambda_{n}) \defeq \sqrt{ \wavenum^2 - \frac{\lambda_n}{r_0^2} } \, ,
\ee
as well as the finite sums (plot~\ref{plot:GlazmanFiniteSums}) of the Glazman condition \eqref{eq:Glazman_condition},
\be
    G_n = \sum_{\substack{i, j = 1 \\ i \ne j}}^{n} 
    \frac{ \Im (\lambda_i)\, \Im (\lambda_j) }{ \vert \lambda_i - \lambda_j \vert^2 }  \,, \quad n=1,2,\ldots
    \label{eq:finite_Glazman}
\ee
For large $r_0$ and $d = 0$, the straight waveguide analysis from~\cite{Demkowicz_Gopalakrishnan_Heuer_24} indicates that 
$$
    -\lambda_n \cong r_0^2 \left( -\wavenum^2 + (n \pi)^2 \right)  \, .
$$
So, for a large bend radius and a low impedance value, one would expect that $n \pi \cong \alpha_n$. 
We use this relation to generate initial guesses for the real component of the eigenvalues. 
After a small pre-asymptotic region, analogous to propagating modes in the straight waveguide, the real part of $\alpha_n$ coincides with $n \pi$ up to the first
five significant digits. At the same time, the imaginary part of $\alpha_n$ converges to zero, asymptotically following the rule $-2bd\kappa/(\pi n)$; cf.~\cite[Lemma 2]{Demkowicz_Gopalakrishnan_Heuer_24}. The third plot displays the finite sums $G_n$ computed by \eqref{eq:finite_Glazman}; here, the sum converges to a finite limit, supporting the hypothesis that condition \eqref{eq:Glazman_condition} holds and, hence, the well-posedness of the problem should be expected.

%

\section{Solver convergence}
\label{sec:convergence}
\revC{This appendix provides additional numerical evidence on the performance of our algorithms implemented in Julia using arbitrary-precision arithmetic.}

\subsection{Power series convergence}
\revC{Algorithm \ref{algo:bessel_frobenius} iterates until the moduli of the last two terms added to the power series \eqref{eq:bessel_frobenius} lie within a certain tolerance. In this subsection, we wish to visualize how the convergence of those power series unfolds. Consider the eigenvalue $\mu=\beta^2/r_0^2$ that was obtained for the first even mode and the PML boundary condition, at different bend radii (see $\beta$ values in the last column of Table \ref{tab:eigval3_comparison_vs_r0}). We evaluate the pair of fundamental solutions $(\Bfirst^{\core}_\mu, \Bsecond^{\core}_\mu)$, and their left and right cladding counterparts, $(\Bfirst^{\lclad}_{\mu_1},\Bsecond^{\lclad}_{\mu_1})$, $(\Bfirst^{\rclad}_{\mu_2},\Bsecond^{\rclad}_{\mu_2})$, at three values of $r$.}

\begin{figure}[htbp]
    \centering
    \begin{subfigure}[b]{0.326\columnwidth}
        \includegraphics[width=\textwidth]{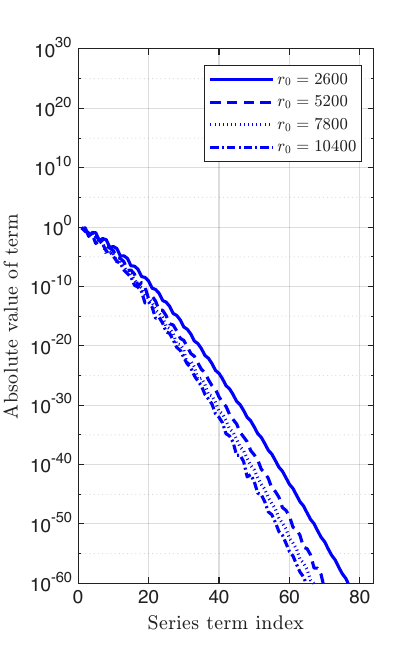}
        \caption{$\Bfirst^{\core}_{\mu}$ evaluated at $r = r_0+a$}
        \label{fig:Vcore-terms-convergence}
    \end{subfigure} \hfill %
    \begin{subfigure}[b]{0.326\columnwidth}
        \includegraphics[width=\textwidth]{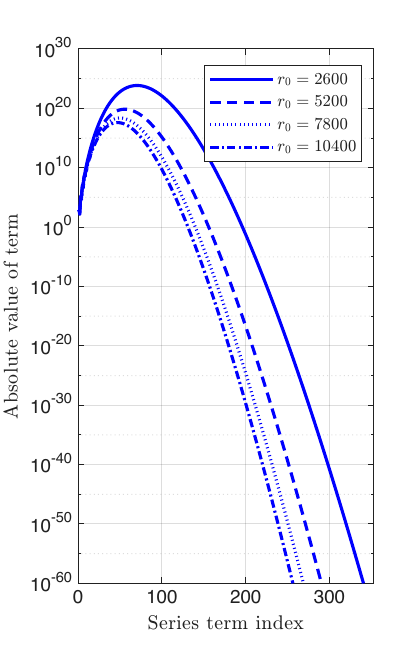}
        \caption{$\Bfirst^{\lclad}_{\mu_1}$ evaluated at $r = r_0-b$}
        \label{fig:Vleft-terms-convergence}
    \end{subfigure} \hfill %
    \begin{subfigure}[b]{0.326\columnwidth}
        \includegraphics[width=\textwidth]{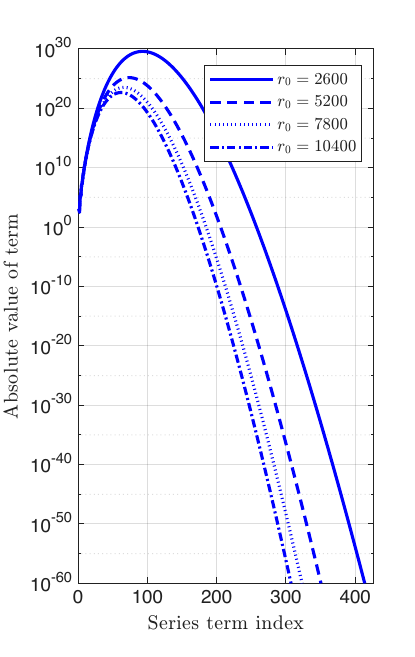}
        \caption{$\Bfirst^{\rclad}_{\mu_2}$ evaluated at $r = \zpml$}
        \label{fig:Vright-terms-convergence}
    \end{subfigure}
    \caption{\revC{Convergence of the rescaled power series for $\Bfirst_{\mu}$, using the computed eigenvalue $\mu$ of the first even mode, evaluated at three relevant points: (a) core-right cladding interface, (b) inner boundary, (c) outer boundary (with PML).}}
    \label{fig:V-terms-convergence}
\end{figure}

\begin{figure}[!htbp]
    \centering
    \begin{subfigure}{0.326\columnwidth}
        \includegraphics[width=\textwidth]{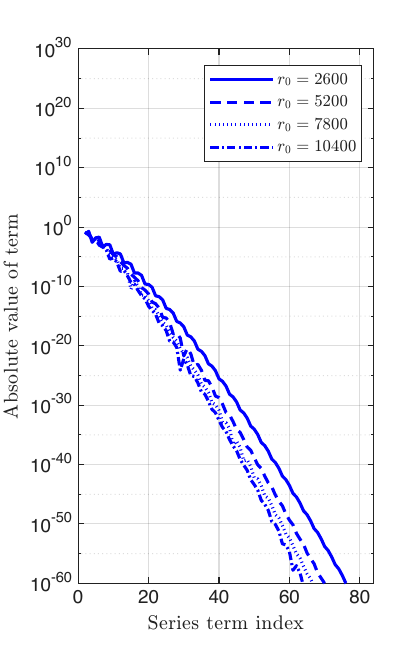}
        \caption{$\Bsecond^{\core}_{\mu}$ evaluated at $r = r_0+a$}
        \label{fig:Wcore-terms-convergence}
    \end{subfigure} \hfill %
    \begin{subfigure}{0.326\columnwidth}
        \includegraphics[width=\textwidth]{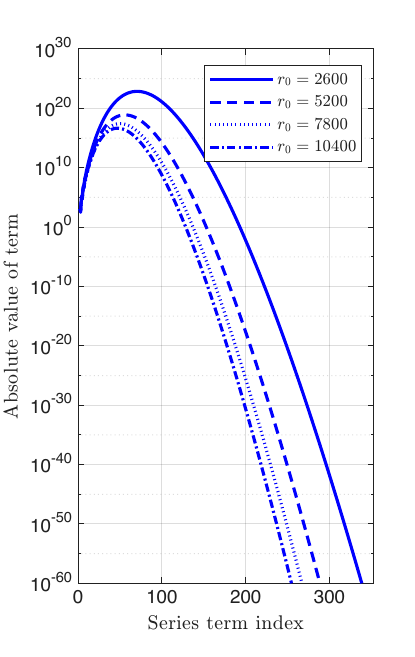}
        \caption{$\Bsecond^{\lclad}_{\mu_1}$ evaluated at $r = r_0-b$}
        \label{fig:Wleft-terms-convergence}
    \end{subfigure} \hfill %
    \begin{subfigure}{0.326\columnwidth}
        \includegraphics[width=\textwidth]{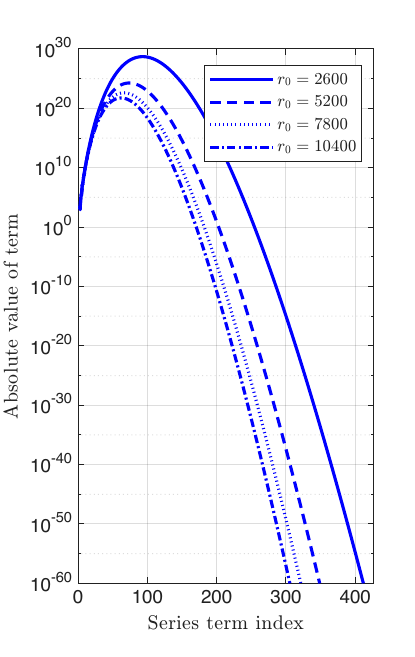}
        \caption{$\Bsecond^{\rclad}_{\mu_2}$ evaluated at $r = \zpml$}
        \label{fig:Wright-terms-convergence}
    \end{subfigure}
    \caption{\revC{Convergence of the rescaled power series for $\Bsecond_{\mu}$, using the computed eigenvalue $\mu$ of the first even mode, evaluated at three relevant points: (a) core-right cladding interface, (b) inner boundary, (c) outer boundary (with PML).}}
    \label{fig:W-terms-convergence}
\end{figure}

\revC{Fig.~\ref{fig:V-terms-convergence} shows the magnitude of every power series term, $|c_m\,x^m|$, for the first fundamental solution, $\Bfirst_\mu$, until convergence is attained, while the data corresponding to the second fundamental solution, $\Bsecond_\mu$, is displayed in Fig.~\ref{fig:W-terms-convergence}. On the left panel of both figures, we see the evaluation of each fundamental solution at point $r=r_0+a$, that is, the right cladding-core interface. It is evident that the terms in this case, for the four values of $r_0$ accounted for, decreased monotonically, and  the desired accuracy was reached in fewer than 80 terms. The middle plot shows the evaluation at $r=r_0-b$ (on the inner boundary). 
As the evaluation point lies farther from the expansion point, the first few dozen series terms initially grow rapidly before decaying, with convergence occurring after approximately 300 terms. 
The right panel in both figures, finally, shows the $r=\zpml$ case (the outer boundary), with an aspect very similar to the previous one, but with a higher maximum and the need for more terms to reach convergence. The reason for these bigger numbers is that $\zpml$ has an imaginary part, which increases the distance from the series' base point. 
Note that, for the second and third plots, it is clear that a double-precision implementation would not have worked properly since the magnitudes of the intermediate series terms grow from approximately $10^0$ to above $10^{20}$ before decreasing toward convergence.}

\subsection{Eigensolver convergence}
\begin{figure}[htbp]
    \centering
    \begin{subfigure}{0.326\columnwidth}
        \includegraphics[width=\textwidth]{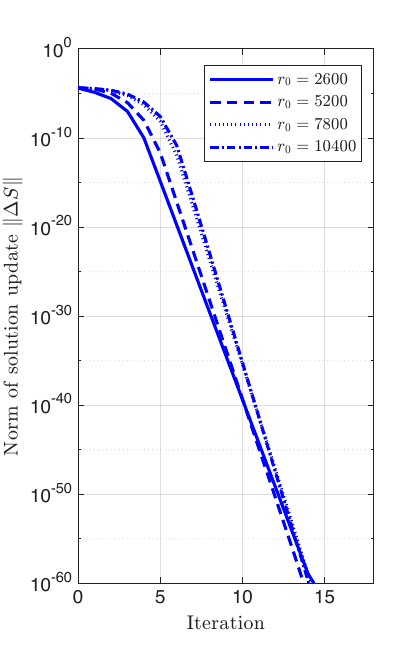}
        \caption{First even mode}
        \label{fig:NR-delta-convergence-even1}
    \end{subfigure} \hfill %
    \begin{subfigure}{0.326\columnwidth}
        \includegraphics[width=\textwidth]{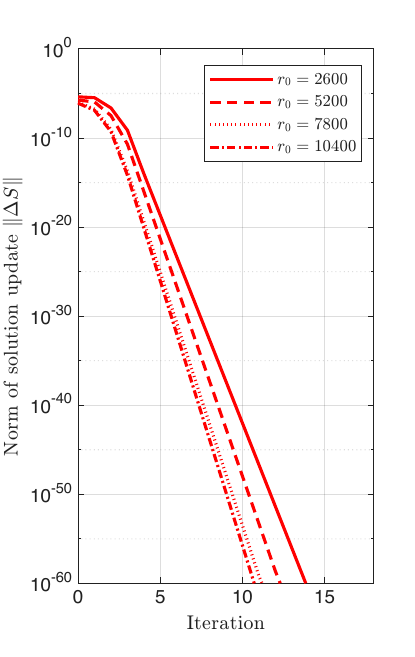}
        \caption{First odd mode}
        \label{fig:NR-delta-convergence-odd1}
    \end{subfigure} \hfill %
    \begin{subfigure}{0.326\columnwidth}
        \includegraphics[width=\textwidth]{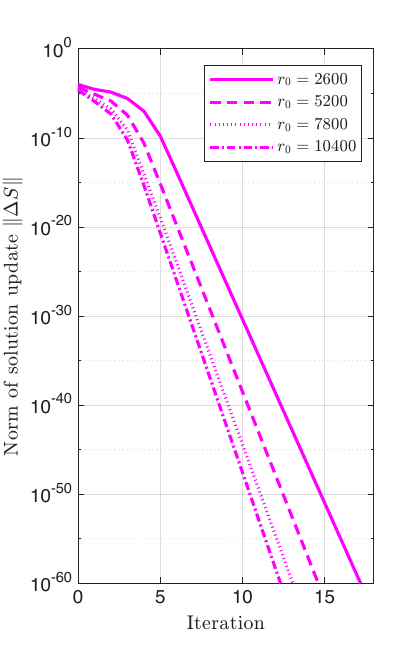}
        \caption{Second even mode}
        \label{fig:NR-delta-convergence-even2}
    \end{subfigure}
    \caption{\revC{Newton--Raphson increment convergence of the eigensolver for three modes at various $r_0$.}}
    \label{fig:NR-delta-convergence}
\end{figure}
\begin{figure}[!htbp]
    \centering
    \begin{subfigure}{0.326\columnwidth}
        \includegraphics[width=\textwidth]{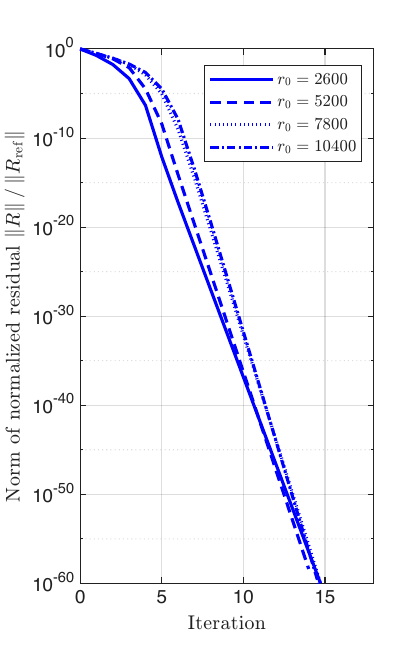}
        \caption{First even mode}
        \label{fig:NR-resid-convergence-even1}
    \end{subfigure} \hfill %
    \begin{subfigure}{0.326\columnwidth}
        \includegraphics[width=\textwidth]{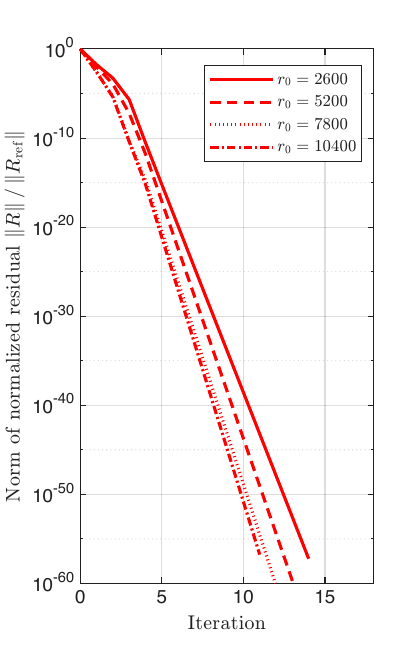}
        \caption{First odd mode}
        \label{fig:NR-resid-convergence-odd1}
    \end{subfigure} \hfill %
    \begin{subfigure}{0.326\columnwidth}
        \includegraphics[width=\textwidth]{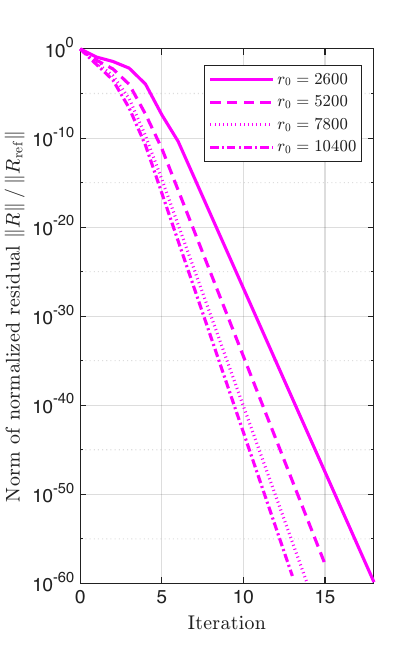}
        \caption{Second even mode}
        \label{fig:NR-resid-convergence-even2}
    \end{subfigure}
    \caption{\revC{Newton--Raphson residual convergence of the eigensolver for three modes at various $r_0$.}}
    \label{fig:NR-resid-convergence}
\end{figure}
\revC{Another iterative technique that supports our methodology is the nonlinear eigensolver, based on a bivariate Newton--Raphson root-finding solver, outlined above in Algorithm~\ref{algo:eigensolver-3layer}. 
To assess the convergence of the PML-based eigensolver for the modes reported in Section~\ref{sec:PML-BC}, we plot the magnitudes of the solution increments and the corresponding nonlinear residuals.
Following the notation introduced in the algorithm pseudocode, the norm of the solution increment (or update) is written as $\|\Delta\bfS\|$, while the norm of the residual is denoted by $\| \bfR \|$. We opt for normalizing the latter, with respect to a reference magnitude, $\|\bfR_{\text{ref}}\|$, which is the norm of the residual evaluated at the initial guess. Fig.~\ref{fig:NR-delta-convergence} presents three plots with the convergence of $\| \Delta\bfS \|$ as the iterations progress, each for a different propagating mode and with four lines representing the chosen bend radii. Similarly, in Fig.~\ref{fig:NR-resid-convergence} the evolution of the normalized residual for those same cases is plotted. Recall that the tolerance of $10^{-60}$ applies only to the solution increment. Indeed, our implementation succeeds in meeting such a threshold: the norm of $\Delta\bfS$ falls below the tolerance, for all the reported cases, in 18 iterations at most. 
Although the residual is not included in the stopping criterion, it decreases to a comparably small magnitude for all reported modes and bend radii, as shown in Fig.~\ref{fig:NR-resid-convergence}.}

\end{document}